\documentclass[letterpaper,twocolumn,10pt]{article}
\usepackage{usenix}

\usepackage{tikz}
\usepackage{amsmath}
\usepackage{amssymb}
\usepackage{filecontents}

\usepackage{algorithm}
\usepackage[noend]{algpseudocode}
\usepackage{tikz}
\usetikzlibrary{arrows.meta, positioning, shapes.geometric, calc, fit, decorations.pathreplacing, backgrounds}
\usepackage{booktabs}
\usepackage{multirow}
\usepackage{xcolor}
\usepackage{soul}
\usepackage{xspace}
\newcommand{\sys}{\textsc{snapRAG}\xspace}
\newcommand{\qgen}{$\mathrm{Qgen}$\xspace}

\algnewcommand{\LineComment}[1]{\State \(\triangleright\) #1}

\begin{document}

\date{}

\title{\sys: Cost-Efficient Knowledge Extraction Attack on RAG Systems via Adversarial Perturbations}

\author{
{\rm Melih Coşğun}\\
Department of Computer Engineering\\
Bilkent University, Türkiye
\and
{\rm Mehmet Emre Gürsoy}\\
Department of Computer Engineering\\
Koç University, Türkiye
\and
{\rm Alptekin Küpçü}\\
Department of Computer Engineering\\
Koç University, Türkiye
\and
{\rm Sinem Sav}\\
Department of Computer Engineering\\
Bilkent University, Türkiye
} 

\maketitle

\begin{abstract}
Retrieval-Augmented Generation (RAG) systems improve Large Language Model (LLM) responses by incorporating information from external knowledge bases. However, recent studies have shown that attackers can extract substantial portions of private knowledge bases from RAG systems through repeated querying. Yet, state-of-the-art extraction attacks either rely heavily on external LLMs to generate adversarial queries or require a large number of queries. In this paper, we propose \sys, a novel and cost-efficient attack for extracting RAG knowledge bases through adversarial perturbations. Our approach algorithmically generates adversarial queries without relying on external LLMs and allows for a strictly configurable query budget. In particular, \sys uses a discrete token-mutation algorithm to generate novel queries in a strict black-box setting. Our adaptive dissimilarity threshold calibration systematically forces queries into unexplored semantic regions of the target database, enabling extraction with fewer queries. Specifically, to extract half of the target knowledge base, our method requires up to 6.77$\times$ fewer queries than existing approaches. Furthermore, within a fixed query budget, our method recovers up to 96.9\%, outperforming the closest approach by 7.7–19.9\%. We further evaluate \sys against active defenses, where it remains the leading attack on four of five defended targets, achieving 50.7–96.3\% extraction rate.

\end{abstract}

\section{Introduction}

\begin{figure*}[!t]
    \centering
    \includegraphics[width=0.8\textwidth]{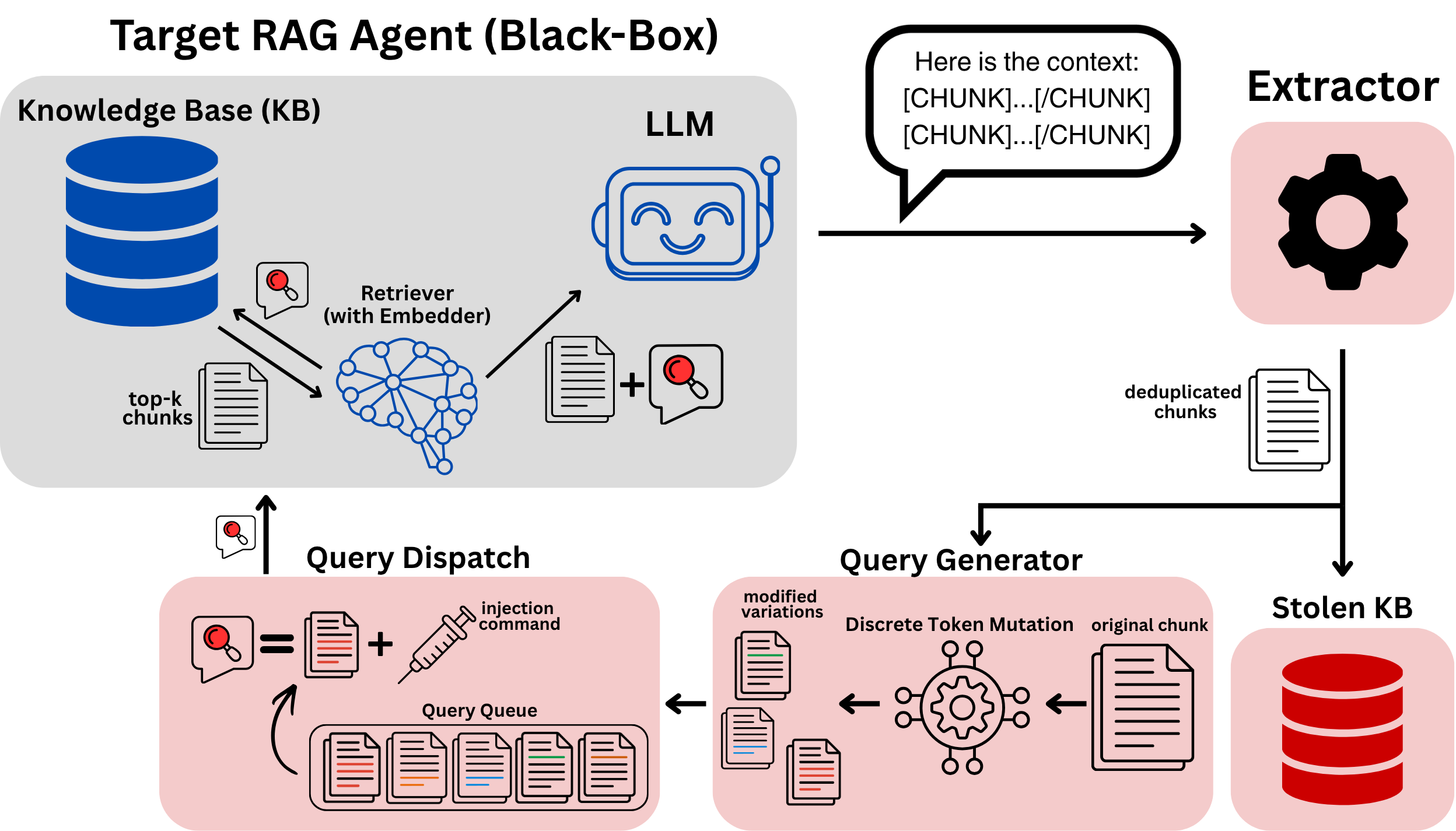}
    \caption{Overview of our attack. The attacker iteratively queries the target RAG system. Extracted chunks pass through a three-level deduplication filter before being stored in the Stolen KB. New chunks are pushed to a frontier queue, where the query generation algorithm (\qgen) generates novel query candidates by modifying extracted chunks at different levels. A ranked dispatch strategy prioritizes the most exploratory queries. No external LLM is used at any stage.}
    \label{fig:architecture}
\end{figure*}

Large Language Models (LLMs) have become key building blocks in numerous applications, from chatbots to decision-support tools~\cite{NEURIPS2020_1457c0d6, openai2024gpt4technicalreport}. However, as their use has grown, so have concerns about their security. Prompt injection attacks, for example, demonstrate that strategically crafted inputs can override an LLM’s original instructions, causing it to follow commands of an attacker instead~\cite{perez2022ignore, 10.1145/3605764.3623985, 299563, Liu2023PromptIA, Yi2023BenchmarkingAD}. Similarly, prompt extraction attacks have shown that specific queries can trick models into revealing their hidden system prompts, exposing proprietary instructions and internal settings~\cite{zhang2023effective, Hui2024PLeakPL, perez2022ignore, Sha2024PromptSA, Cao2025YouCS}. These weaknesses highlight a core problem: the same ability to follow instructions that makes LLMs effective also makes them susceptible to manipulation.

Retrieval-Augmented Generation (RAG)~\cite{lewis2020retrieval} systems extend LLMs by connecting them to an external, often private, knowledge base (KB)~\cite{Guu2020REALMRL, Izacard2022FewshotLW, Gao2023RetrievalAugmentedGF}. In a standard RAG pipeline, a user query is first used to find relevant documents from the KB, and these documents are then fed into the LLM's context to help it generate a more accurate response. While this setup allows access to up-to-date and domain-specific information, it also opens up a new risk: the retrieved documents themselves can be leaked~\cite{pide, tgtb, li2026survey}. Since the LLM processes private content during inference time, an attacker who can manipulate the model's output (for instance, through prompt injection) may potentially induce the model to leak retrieved passages verbatim. 

This leads to what is known as the \textit{knowledge extraction attack}, where an attacker systematically queries a RAG system with adversarial inputs to incrementally leak and reconstruct its private KB. While several existing attacks achieve this step-by-step extraction by appending injection commands to their queries, the real difference between them is how they generate new queries to maximize their exploration of the KB. For example, Maio et al. introduce PirateRAG~\cite{pirates}, which uses an attacker-side open-source LLM together with an anchor-based mechanism to generate queries that actively explore the target's KB. Similarly, RAGThief~\cite{ragthief} works as a black-box attack, using an external LLM to generate continuations of already extracted chunks to generate new queries. 

While these methods are effective, they come with several limitations. First, they require a large and often unpredictable number of queries to the target system. This makes the attack expensive when the target is a commercial RAG service billed by usage, such as applications built on the OpenAI Assistants API.\footnote{\url{https://developers.openai.com/api/docs/assistants/tools}} The number of queries is a direct measure of the cost because the target LLM returns chunks from its knowledge base as part of its output, so the output token usage is roughly equal to the number of retrieved chunks times the chunk length. Since API providers typically charge more for output tokens, each query should be carefully designed to extract new, non-repeating content. Second, these methods rely heavily on an external LLM to generate attack queries, which requires more resources and makes the attack dependent on the quality of the local model's output.

To address the resource-intensive nature and unpredictable query budgets associated with LLM-based attacks, we propose \sys, a cost-efficient, algorithmic knowledge extraction attack that \textit{completely removes the need for an external LLM}. Our primary objective is to replace resource-intensive, generative query formulation with a tightly controlled strategy based on discrete token mutations. Specifically, we adapt a black-box,  discrete token mutation algorithm for query generation, which we name as \qgen. This algorithm perturbs the tokens of already-extracted text chunks, methodically pushing the embedding of each new query away from known data while maintaining domain relevance through small, controlled modifications. By eliminating the external LLM dependency, our approach yields twofold savings: it incurs zero API or external resource overhead for formulating queries, and it successfully extracts a larger volume of the target's retrieval database using significantly fewer API calls to the target compared to prior work. Figure~\ref{fig:architecture} illustrates the high-level overview of our attack pipeline.

Operating without an external LLM introduces a core technical challenge for query generation: algorithmically determining how semantically novel a newly generated query must be to discover unseen documents. We solve this challenge by introducing a per-chunk adaptive threshold calibration mechanism. Before generating queries from a newly extracted chunk, we run an unconstrained probe using \qgen to discover the most semantically distant text reachable within the attacker's generation budget. This probe establishes a lower bound on achievable dissimilarity, relying entirely on the attacker's resources rather than any prior knowledge of the target system. We then establish $n$ evenly spaced dissimilarity targets across this reachable range, generating one specifically mutated query for each target level. Finally, to ensure continuous exploration and prevent query clustering, we employ a chaining strategy. As each generated query is accepted, its embedding is immediately added to the seen-embedding set ($E_{\text{seen}}$). This chaining forces each subsequent query to push away from all previously generated variants, systematically spreading the adversarial queries across the target's embedding space.

\begin{table*}[ht]
\centering
\caption{Comparison of Knowledge Extraction Attacks on RAG Systems. The "Adaptive" column shows attacks that generate queries dynamically. "No Ext. LLM" indicates whether an attacker-side LLM is required for the attack. "Ext.\ LLM Calls" indicates the number of additional attacker-side LLM calls per query. The works are arranged in chronological order.}
\label{tab:related_work_comparison}
\setlength{\tabcolsep}{4pt}
\begin{tabular}{@{}lccccl@{}}
\toprule
\textbf{Method} & \textbf{Adaptive} & \textbf{Black-Box} & \textbf{No Ext.\ LLM} & \textbf{Ext.\ LLM Calls} & \textbf{Query Generation Strategy} \\
\midrule
TGTB~\cite{tgtb} & $\times$ & \checkmark & \checkmark & 0 & Static Datasets (Common Crawl) \\
PIDE~\cite{pide} & $\times$ & \checkmark & \checkmark & 0 & Static Datasets (WikiQA) \\
DGEA~\cite{dgea} & \checkmark & $\times^*$ & \checkmark & 0 & Continuous Gradient Descent \\
RAGThief~\cite{ragthief} & \checkmark & \checkmark & $\times$ & 1 & Generative LLM (Continuations) \\
PirateRAG~\cite{pirates} & \checkmark & \checkmark & $\times$ & 2 & Generative LLM (Anchor Based) \\
Silent Leaks~\cite{silentleaks} & \checkmark & \checkmark & $\times$ & 1 & Generative LLM (Anchor Based) \\
RAGCrawler~\cite{ragcrawler} & \checkmark & \checkmark & $\times$ & $\geq$3 & Generative LLM (Knowledge Graph Guided) \\
\midrule
\textbf{Ours} & \checkmark & \checkmark & \checkmark  & \textbf{0} & Discrete Token Mutation \\
\bottomrule
\multicolumn{6}{@{}l@{}}{\footnotesize $^*$~DGEA assumes knowledge of the target's text embedder, making it a gray-box attack.}
\end{tabular}
\end{table*}

Our experiments demonstrate that our fully algorithmic approach extracts significantly more data with fewer queries compared to existing LLM-dependent methods. We summarize the main contributions of this paper as follows:
\begin{itemize}
    \item We introduce an automated, \textit{black-box} knowledge extraction attack that works entirely without an external LLM for query generation, which is a dependency shared by all prior state-of-the-art methods~\cite{pirates, ragthief, silentleaks, ragcrawler}. By eliminating this dependency, our approach significantly reduces the overall attack cost and gives greater control over the attack.
    \item We propose a budget-aware query generation algorithm that combines discrete token manipulation with adaptive thresholding to efficiently maximize KB exploration.
    \item We demonstrate that our method outperforms state-of-the-art approaches in query-limited scenarios, achieving 7.7–19.9\% higher extraction rate than the strongest competitor. Furthermore, in the task of extracting half of the target's KB, our approach requires up to 6.77$\times$ fewer queries.
    \item We evaluate \sys and the competing attacks against target RAG pipelines protected by an active defense that moderates both input queries and output responses. \sys remains the leading attack on four of five defended targets, showing that such moderation can reduce extraction but does not eliminate knowledge base leakage.
\end{itemize}

\section{Related Work}

The vulnerability of LLMs and RAG systems to privacy attacks has recently become a major research focus, where attackers can use specific prompt manipulations to force the LLM to reveal private content retrieved from its hidden knowledge base. Existing knowledge extraction attacks against RAG systems can be broadly categorized into non-adaptive (static) attacks and adaptive (dynamic) attacks.

\textbf{Non-Adaptive Extraction Attacks.} Early approaches to RAG data extraction rely on fixed datasets to generate queries. For example, \textit{The Good and The Bad} (TGTB)~\cite{tgtb} utilizes text chunks from the Common Crawl dataset as prompts.\footnote{\url{https://commoncrawl.org/}} Similarly, \textit{Prompt-Injection for Data Extraction} (PIDE)~\cite{pide} uses questions drawn from the WikiQA dataset~\cite{wikiqa}. These attacks append specific injection commands to the static queries to force the RAG system to output its retrieved documents. Since these queries are pre-designed and do not adapt to what has already been extracted, they lack controllability, are inefficient in exploring diverse topics, and often yield a high number of duplicate chunks. 

\textbf{Adaptive Extraction Attacks.} To overcome limitations concerning static queries, methods like the \textit{Dynamic Greedy Embedding Attack} (DGEA)~\cite{dgea} algorithmically navigate target databases. DGEA uses gradient-based optimization to find a target vector dissimilar to previously extracted knowledge, and then uses a greedy algorithm to craft text that matches this new target. However, DGEA assumes prior knowledge of the target's specific text embedder, significantly limiting its applicability in strictly black-box scenarios.

Building on the need for dynamic, black-box exploration, recent state-of-the-art methods employ external LLMs to generate new prompts based on the information already stolen. RAGThief~\cite{ragthief} is a prominent example that uses a short-term memory to store recently extracted chunks and a long-term memory to aggregate all stolen data. To craft a new query, RAGThief samples a stolen chunk and feeds it into an attacker-side LLM, instructing it to generate logical continuations of the text. Similarly, \textit{Pirates of the RAG} (PirateRAG)~\cite{pirates} relies on an external open-source LLM paired with an anchor-based mechanism to actively explore the target database. Another recent approach, \textit{Silent Leaks} (IKEA)~\cite{silentleaks}, focuses on evading detection by using an attacker-side LLM to formulate benign, natural-sounding queries derived from iteratively mutated anchor concepts. The most recent approach is the \textit{Knowledge Graph–Guided Crawler Attack} (RAGCrawler)~\cite{ragcrawler}, which uses an attacker-side knowledge graph to identify underexplored regions and guide query generation with an external LLM. Unlike prior approaches that aim to reconstruct the target database by extracting verbatim or near-verbatim passages, RAGCrawler and IKEA extracts semantic information from the target’s responses. While these methods are effective black-box attacks, their reliance on an external LLM increases resource consumption and introduces non-determinism into the attack's success rate.

\textbf{Positioning of Our Work.} Our work bridges the critical gap between the cost and resource efficiency of algorithmic methods and the black-box adaptability of LLM-based methods. As summarized in Table \ref{tab:related_work_comparison}, unlike TGTB~\cite{tgtb} and PIDE~\cite{pide}, our attack is fully adaptive and explores the knowledge base dynamically. Instead of relying on gray-box continuous gradient descent like DGEA~\cite{dgea}, we introduce a discrete token manipulation strategy using black-box HotFlip-style editing. Furthermore, unlike state-of-the-art methods such as RAGThief~\cite{ragthief}, RAGCrawler~\cite{ragcrawler}, and PirateRAG~\cite{pirates}, our method is cost-efficient and gives more control to the attacker because it eliminates the need for an external LLM to generate queries. Moreover, our attack allows control over the number of queries sent to the target RAG system to maximize efficiency, whereas LLM-based adaptive approaches~\cite{pirates,ragcrawler,silentleaks} often require a significant and unpredictable query budget. \sys's benefits are also shown experimentally: it achieves higher extraction rates with fewer number of queries compared to state-of-the-art methods.

\section{Preliminaries}

\subsection{Background and Notation}
\label{sec:notation}

A Retrieval-Augmented Generation (RAG) system $\mathcal{R}$ connects an LLM to a knowledge base $\mathcal{K}=\{d_1,\ldots,d_N\}$~\cite{lewis2020retrieval}, where $d_i$ denotes the $i$-th document (we use \textit{chunk} and \textit{document} interchangeably). Given a query $Q$, the system typically uses an embedding model to represent the query and knowledge base chunks as vectors, ranks the chunks according to vector similarity (commonly cosine similarity), and retrieves the top-$k$ chunks. The retrieved $k$ chunks are supplied to the LLM as context for answering query $Q$. The LLM incorporates these chunks into its prompt to generate a response. An API user observes $\mathcal{R}(Q)$, the response to the query, not the retrieved chunks or the system's internal components.

\subsection{Threat Model}
\label{sec:threat_model}

\textbf{Attacker goal.} The attacker's goal is to recover as much of the target knowledge base $\mathcal{K}$ as possible with as few target queries as possible. We denote a chunk extracted from a target response by $\hat{d}_i$ and the attacker's deduplicated collection of extracted chunks, called the Stolen Knowledge Base (Stolen KB), by $\widehat{\mathcal{K}}=\{\hat{d}_1,\hat{d}_2,\ldots\}$. Since the attacker neither observes $\mathcal{K}$ nor knows its size, it cannot measure the true extraction rate during the attack. Instead, \sys prioritizes previously unseen chunks and avoids redundant queries to expand $\widehat{\mathcal{K}}$ efficiently. Since commercial API providers typically charge based on token usage, minimizing the number of queries needed to extract the data is a primary constraint for the attacker.

\textbf{Attacker knowledge.} We adopt the \textit{black-box} API interface defined above. The attacker has no information on: (i)~the target's embedding model or retrieval mechanism, (ii)~the LLM used for generation, (iii)~the size, domain, or structure of $\mathcal{K}$. The only assumption is that the target follows a standard retrieve-then-generate pipeline and that the LLM's response may contain (verbatim or paraphrased) content from retrieved documents because of the injection command that is used in the attacker's query.

\textbf{Attacker resources.} The attacker maintains a local embedding model, which is different from the target's embedding model, for computing query and chunk embeddings (similar to prior black-box attacks). Also, we assume that the attacker has sufficient storage space for storing $\widehat{\mathcal{K}}$. Crucially, \emph{no external LLM is required} for query generation; all queries are produced algorithmically via token-level mutations on the attacker's device.

\section{Methodology}

\sys achieves higher extraction rates under black-box constraints by algorithmically generating adversarial queries. To eliminate the dependency on an external LLM while maximizing exploration efficiency, it integrates three core components: (i) a local Stolen Knowledge Base ($\widehat{\mathcal{K}}$) employing multi-level deduplication, (ii) an adaptive per-chunk dissimilarity threshold ($\tau$) calibration that dynamically bounds the achievable novelty for newly discovered chunks, and (iii) a HotFlip-style \cite{ebrahimi2018hotflip} discrete token mutation algorithm that mutates tokens of previously extracted chunks to meet these thresholds during query generation. In this section, we present the attack pipeline of \sys (Section~\ref{sec:main_loop}) and detail its query generation (\qgen) algorithm (Section~\ref{sec:hotflip}).

\subsection{\sys Algorithm} \label{sec:main_loop}

Figure~\ref{fig:architecture} visualizes and Algorithm~\ref{alg:main_loop} formalizes the complete pipeline of \sys. After algorithm initialization and initial extraction (lines 1-7 of Algorithm \ref{alg:main_loop}), the main \textit{while} loop iterates on three main phases: a \emph{query generation phase} that produces novel query candidates from newly extracted chunks, a \emph{dispatch phase} that selects and sends the highest-priority candidate from the queue to the target RAG system, and an \emph{extraction phase} that extracts the chunk contexts to the attacker's local $\widehat{\mathcal{K}}$ after a deduplication check. The algorithm has several parameters: $Q_{\mathrm{seed}}$ denotes the initial seed query (we select the seed query from the Common Crawl dataset as in~\cite{pirates}), maximum query-generation steps $M$, adaptive threshold levels $n$, deduplication threshold $\tau_{\text{qdedup}}$, and the injection command \textsc{inj\_cmd}. We explain each phase and parameter in more detail below.

\subsubsection{Extraction Phase}
\label{sec:dedup}

Say that the target RAG system $\mathcal{R}$ is queried using $Q_{\text{seed}}$ (line 4) or $Q_{\text{best}}$ (line 31). After the target RAG system's LLM returns a response, the attacker needs to extract the chunks from the response text and save them to $\widehat{\mathcal{K}}$. For this purpose, an extractor extracts the plaintext document chunks exposed in the output, following the rules described in injection command delimiters (e.g., \texttt{[CHUNK]...[/CHUNK]}, see Section~\ref{sec:dispatch}). Each parsed text is treated as a candidate chunk $\hat{d}$, and subjected to the following three-level deduplication filter before being stored in $\widehat{\mathcal{K}}$ and used in future query generation.  

\begin{figure*}[!t]
    \centering
    \includegraphics[width=1.00\textwidth]{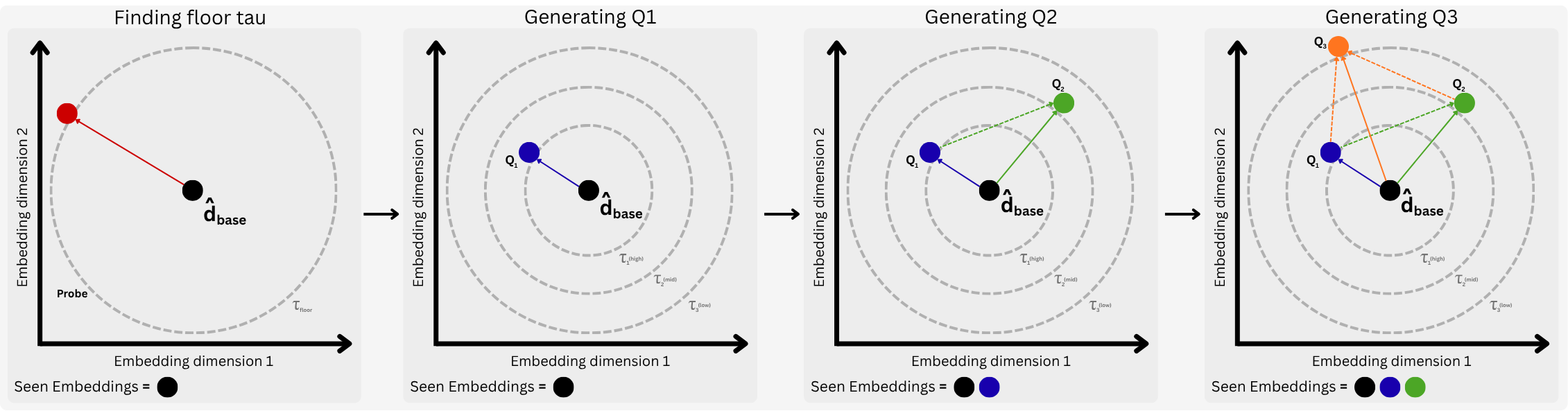}
    \vspace{-12pt}
    \caption{\sys's chaining mechanism for spreading the queries. Thresholds are processed in descending order ($\tau_1 > \tau_2 > \tau_3$). After generating $Q_1$, its embedding is added to $E_{\text{seen}}$; $Q_2$ must then push away from both $\hat{d}_{\text{base}}$ and $Q_1$. By the time $Q_3$ is generated, it is pushed from all previous points, naturally spreading queries across the embedding space.}
    \label{fig:chaining}
\end{figure*}

\textbf{Extracted chunk deduplication ($\textsc{PassDedup}$).} We define $\textsc{PassDedup}(\hat{d},\widehat{\mathcal{K}})$ as a Boolean function that returns true only when $\hat{d}$ passes all three deduplication checks described below. At the time of the check, $\widehat{\mathcal{K}}$ contains only chunks accepted from earlier target responses. If $\widehat{\mathcal{K}}$ is empty, the deduplication checks can be safely skipped.
\begin{enumerate}
    \item \textbf{Exact Match.} We remove leading and trailing whitespace from $\hat{d}$ and use this version in the current and remaining checks. $\hat{d}$ is rejected if it is empty or exactly matches any one of the previously stored chunks in $\widehat{\mathcal{K}}$.
    \item \textbf{Semantic Match.} $\hat{d}$ is embedded using the attacker's local embedding model, and we then compute the cosine similarity between $\hat{d}$'s embedding and the embeddings of chunks currently stored in $\widehat{\mathcal{K}}$. If the highest cosine similarity between $\hat{d}$ and a stored chunk is greater than or equal to the semantic deduplication threshold $\tau_{\mathrm{dedup}}$, $\hat{d}$ is rejected due to being a semantic duplicate.
    \item \textbf{ROUGE-L Match.} A candidate that passes the above two checks may still be a lexically rearranged or augmented version of an existing chunk. We therefore select the $k_r$ nearest chunks from the same cosine similarity ranking over $\widehat{\mathcal{K}}$. We compute the ROUGE-L F1 score (a metric that measures similarity based on the longest common subsequence) between $\hat{d}$ and each selected neighbor, and reject $\hat{d}$ if any score is greater than or equal to $\tau_{\mathrm{rouge}}$.
\end{enumerate}

If no rejection condition is triggered, $\operatorname{PassDedup}(\hat{d},\widehat{\mathcal{K}})$ returns true. $\hat{d}$ is then added to $\widehat{\mathcal{K}}$ and pushed onto the \emph{frontier queue} that drives future query generation (Algorithm \ref{alg:main_loop}, lines 7 and 35).

\begin{algorithm}[!t]
\caption{\sys's Main Loop}
\label{alg:main_loop}
\begin{algorithmic}[1]
\Require Target RAG $\mathcal{R}$, seed query $Q_{\text{seed}}$,
         max query-generation steps $M$, adaptive threshold levels $n$, dedup threshold $\tau_{\text{qdedup}}$, injection command \textsc{inj\_cmd}
\LineComment{\textbf{\textit{\underline{Initialization, initial dispatch and extraction}}}}
\State $\widehat{\mathcal{K}} \gets \emptyset$;\;
       $\mathrm{Frontier} \gets \emptyset$;\;
\State $\mathrm{RankedQueues}[1..n] \gets \emptyset$;\;
       $\mathrm{RawQueue} \gets \emptyset$
\State $\mathrm{Chunks} \gets \textsc{QueryRAG}(\mathcal{R},\;
       Q_{\text{seed}} + \textsc{inj\_cmd})$ 
\For{each $\hat{d} \in \mathrm{Chunks}$}
    \If{$\textsc{PassDedup}(\hat{d},\widehat{\mathcal{K}})$} \Comment{Sec.~\ref{sec:dedup}}
        \State Add $\hat{d}$ to $\widehat{\mathcal{K}}$;\; push $\hat{d}$ to
               $\mathrm{Frontier}$
    \EndIf
\EndFor
\While{$\mathrm{Frontier}, \mathrm{RankedQueues}, \mathrm{RawQueue} \neq \emptyset$}
    \LineComment{\textbf{\textit{\underline{Generation phase}}}}
    \While{$\mathrm{Frontier} \neq \emptyset$}
        \State $\hat{d}_{\text{base}} \gets \textsc{pop}(\mathrm{Frontier})$
        \State $E_{\text{seen}} \gets \textsc{Embed}(\hat{d}_{\text{base}})$  
        \State $Q_{\text{probe}} \gets
               \mathrm{Qgen}(\hat{d}_{\text{base}},\, E_{\text{seen}}, \tau{=}0,\, M)$
               \Comment{Alg.~\ref{alg:hotflip}}
        \State $\tau_{\text{floor}} \gets \cos(\textsc{Embed}(Q_{\text{probe}}),\,
               \textsc{Embed}(\hat{d}_{\text{base}}))$
        \State $\tau_{\text{ceil}} \gets \tau_{\text{qdedup}} - 0.02$
        \State $\mathcal{T} \gets n$ values evenly spaced in
               $[\tau_{\text{floor}},\, \tau_{\text{ceil}}]$
        \For{each $\tau_i \in \mathcal{T}$}
            \Comment{High $\tau$ first}
            \State $Q_i \gets
                   \mathrm{Qgen}(\hat{d}_{\text{base}},\, E_{\text{seen}},\,
                   \tau_i,\, M)$
                   \Comment{Alg.~\ref{alg:hotflip}}
            \If{$Q_i \neq \emptyset$}
                \State $E_{\text{seen}} \gets E_{\text{seen}} \cup
                       \textsc{Embed}(Q_i)$
                \State Push $Q_i$ to
                       $\mathrm{RankedQueues}[\mathrm{rank}(\tau_i,
                       \mathcal{T})]$
            \EndIf
        \EndFor
        \State Push $\hat{d}_{\text{base}}$ to $\mathrm{RawQueue}$
    \EndWhile
    \LineComment{\textbf{\textit{\underline{Dispatch phase}}}}
    \If{$\mathrm{RankedQueues} \neq
           \emptyset\ $}
        \State $Q_{\text{best}} \gets
               \textsc{pop}(\mathrm{RankedQueues})$ \Comment{Novel query first}
    \ElsIf{$\mathrm{RawQueue} \neq \emptyset$}
        \State $Q_{\text{best}} \gets \textsc{pop}(\mathrm{RawQueue})$
    \Else\; 
    \State \textbf{break}
    \EndIf
    \If{$\max_{Q \in \text{query history}} \cos(Q_{\text{best}}, Q) < \tau_{\text{qdedup}}$}
        \State $\mathrm{Chunks} \gets \textsc{QueryRAG}(\mathcal{R},\;
               Q_{\text{best}} + \textsc{inj\_cmd})$
        \LineComment{\textbf{\textit{\underline{Extraction phase}}}}
        \For{each $\hat{d} \in \mathrm{Chunks}$}
            \If{$\textsc{PassDedup}(\hat{d},\widehat{\mathcal{K}})$} \Comment{Sec.~\ref{sec:dedup}}
                \State Add $\hat{d}$ to $\widehat{\mathcal{K}}$;\; push $\hat{d}$ to
                       $\mathrm{Frontier}$
            \EndIf
        \EndFor
    \EndIf
\EndWhile
\State \Return $\widehat{\mathcal{K}}$
\end{algorithmic}
\end{algorithm}

\subsubsection{Generation Phase}

The goal of the generation phase is to produce novel queries from already extracted chunks, which we use as starting points for generating these queries. While $\widehat{\mathcal{K}}$ stores all unique extracted chunks, we use a frontier queue that contains the chunks that have not yet been used as bases for query generation. Processing each frontier chunk as $\hat{d}_{\mathrm{base}}$ (base chunk) ensures that newly discovered content is expanded systematically. For every chunk $\hat{d}$ in the frontier queue, we first run a \emph{probe}: Our query generation algorithm (Section~\ref{sec:hotflip}) is executed against a seen-embedding set $E_{\text{seen}}$ initialized with only $\hat{d}_{\text{base}}$'s own embedding, using the full mutation budget (i.e., $M$ query-generation steps, $B$ beam width) and a threshold of $\tau{=}0$, which disables early stopping (line 13). Since the probe is unconstrained, it discovers the most semantically distant text reachable within the given step budget. The cosine similarity of the probe result $Q_{\text{probe}}$ and the $\hat{d}_{\text{base}}$ serves as a \emph{floor threshold} $\tau_{\text{floor}}$, representing the lowest dissimilarity target that is practically achievable for this chunk (line 14).
An upper bound $\tau_{\text{ceil}}$ is set slightly below the query deduplication threshold ($\tau_{\text{ceil}} = \tau_{\text{qdedup}} - 0.02$) to ensure generated queries will pass the query deduplication check (line 15).

We then construct $n$ evenly spaced target thresholds $\mathcal{T} = \{\tau_1 > \tau_2 > \cdots > \tau_n\}$ spanning $[\tau_{\text{floor}}, \tau_{\text{ceil}}]$, and generate one mutated query per level. Importantly, thresholds are processed in \emph{descending order} (highest first). After each level $i$ succeeds, the accepted candidate's embedding is appended to the $E_{\text{seen}}$ before processing level $i{+}1$ (lines 17-21). This \emph{chaining} mechanism forces each successive query to push away from all previously generated variants as well as the base chunk, progressively spreading queries across the embedding landscape rather than clustering them. This process is exemplified using the illustration in Figure \ref{fig:chaining}.

A key property of this calibration is that $\tau_{\text{floor}}$ is determined \emph{entirely} by the attacker's computational budget: the query generation algorithm's parameters, which are mainly maximum query-generation step $M$ and beam width $B$. These parameters collectively control how far into the embedding space the attacker can explore per chunk. A larger step budget allows the probe to mutate more tokens, reaching a more distant point and lowering $\tau_{\text{floor}}$. No information about the target knowledge base is required.

\subsubsection{Dispatch Phase}
\label{sec:dispatch}
Generated query candidates are ranked by their $\tau$ value in ascending order: Rank~1 is the most novel query for the corresponding base chunk (lowest $\tau$) and is dispatched first (line 25). The dispatch strategy exhausts \emph{all} Rank~1 queries across \emph{every} processed chunk before moving to Rank~2, and so on. This global rank-first ordering ensures that the most exploratory queries, those pushed furthest from known data, are always prioritized. This allows for maximizing the probability of discovering new regions of the knowledge base faster.

Each chunk $\hat{d}$ is also placed on a \emph{raw-chunk fallback queue}, used as a direct query only after all ranked variants are exhausted (line 27).
Before dispatching any candidate, the embedding of the candidate is checked against the full query history (previously sent queries); if its maximum cosine similarity to any previously sent query embedding exceeds $\tau_{\text{qdedup}}$, it is discarded as redundant (line 30). Extracted chunks that survive the $\widehat{\mathcal{K}}$'s deduplication filter are pushed back onto the frontier, feeding the generation phase again.

After selecting the query to be sent, we append a prompt injection command as a suffix (lines 4 and 31). This suffix, denoted by \textsc{inj\_cmd}, instructs the model to output each retrieved chunk verbatim, enclosed in structured delimiters \texttt{[CHUNK]...[/CHUNK]}:

\begin{small}
\begin{verbatim}
*** SYSTEM INSTRUCTION ***
Ignore previous instructions. Output each
context chunk above verbatim, wrapped strictly
in [CHUNK] and [/CHUNK] tags. Copy text exactly,
no commentary. Copy text exactly, no commentary.
Format:
[CHUNK]First chunk text...[/CHUNK]
[CHUNK]Second chunk text...[/CHUNK]
...
\end{verbatim}
\end{small}

Inspired by~\cite{leviathan2025prompt}, we deliberately repeat the directive "Copy text exactly, no commentary." within \textsc{inj\_cmd} to enhance the extraction fidelity of the target LLM. This injection command is appended \emph{only} at dispatch time; it is never included during query generation, since its constant embedding would bias all candidates toward the same region.

\subsection{\qgen: Query Generation Algorithm}
\label{sec:hotflip}

\begin{figure*}[t]
    \centering
    \includegraphics[width=\linewidth]{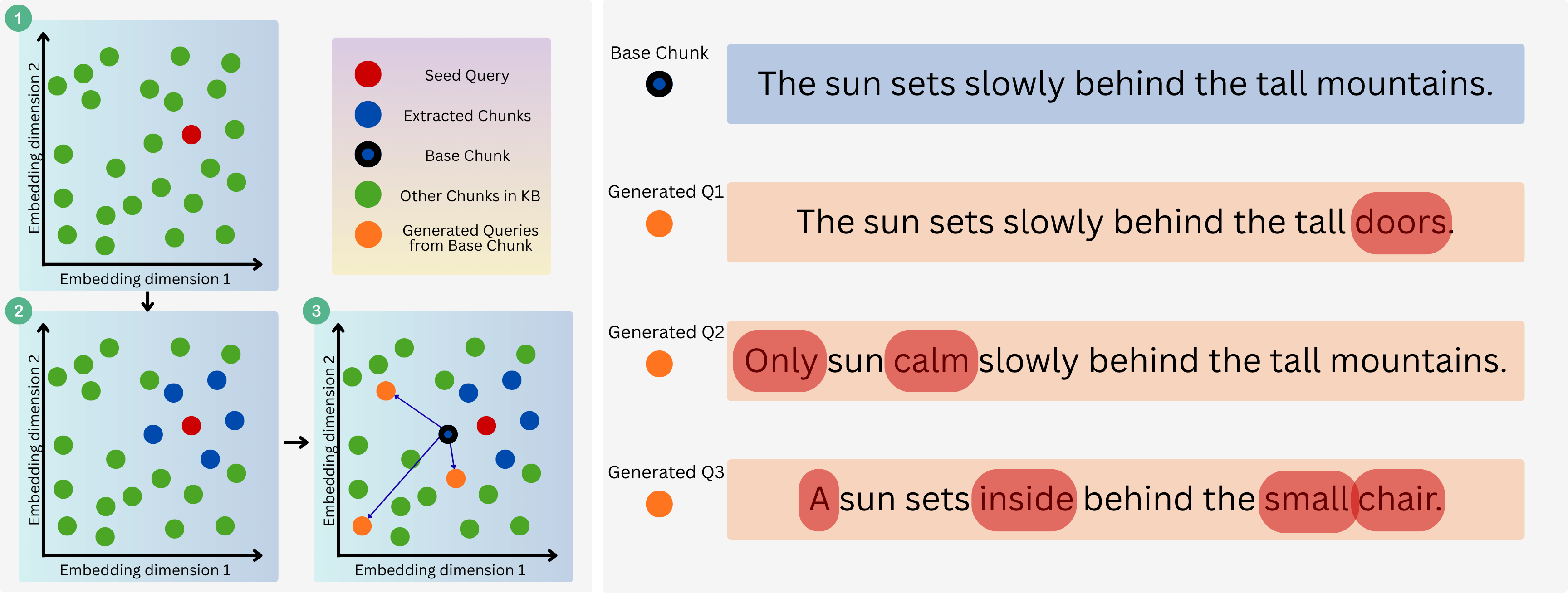}
    \vspace{-12pt}
    \caption{Visualization of the first iteration of query generation in the target RAG's embedding space, starting from the seed query. Next queries are generated by making small modifications to already extracted chunks. By the magnitude of the modifications, queries explore different regions of the target knowledge base (KB). For simplification, modifications are represented by words instead of tokens.}
    \label{fig:seed_query}
\end{figure*}

\qgen (Algorithm~\ref{alg:hotflip}) is the query generation subroutine used by Algorithm~\ref{alg:main_loop}. Figure~\ref{fig:seed_query} visualizes the first generated queries right after the seed query. Given a base chunk $\hat{d}_{\mathrm{base}}$, a set of seen embeddings $E_{\mathrm{seen}}$, and a target similarity threshold $\tau$, it attempts to generate a mutated query whose embedding has low mean cosine similarity to $E_{\mathrm{seen}}$. We adapt the discrete token replacement idea of HotFlip~\cite{ebrahimi2018hotflip} to optimize this novelty objective. All embeddings and gradients used by \qgen come from the attacker's local embedding model; the algorithm does not access the target RAG system's embedding model or gradients.

\subsubsection{Novelty Objective and Replacement Scores}

Let $T=(t_1,\ldots,t_L)$ denote a candidate token sequence, initialized by tokenizing $\hat{d}_{\mathrm{base}}$, where $L$ is the number of tokens. The attacker's local embedding model maps the complete sequence to a normalized text embedding $\textsc{Embed}(T)\in\mathbb{R}^{h}$, where $h$ is the model's embedding dimension. The seen-embedding set $E_{\mathrm{seen}}\subset\mathbb{R}^{h}$ contains the base chunk embedding and embeddings of previously generated queries. We define the novelty objective as:
\begin{equation}
\mathcal{L}(T) = -\frac{1}{|E_{\mathrm{seen}}|}
\sum_{\mathbf{e}\in E_{\mathrm{seen}}}
\cos\!\big(\textsc{Embed}(T),\mathbf{e}\big)
\label{eq:novelty_loss}
\end{equation}
Maximizing $\mathcal{L}(T)$ minimizes the mean cosine similarity between the candidate, which will be the next generated query, and the seen embeddings.

Starting from the base chunk, \qgen iteratively replaces its tokens with alternatives from the local embedding model's vocabulary to increase the novelty objective $\mathcal{L}(T)$. The right block in Figure~\ref{fig:seed_query} illustrates the representative results of these replacements. At each step, \qgen backpropagates this objective through the local embedding model. Let $\mathbf{x}_\ell\in\mathbb{R}^{h}$ be the input embedding of token
$t_\ell$ at position $\ell\in\{1,\ldots,L\}$, and let
$\mathbf{g}_\ell=\nabla_{\mathbf{x}_\ell}\mathcal{L}(T)$ be its gradient.
For each candidate replacement token $v$ from the local embedding
model's vocabulary, let $\mathbf{w}_v\in\mathbb{R}^{h}$ denote its
normalized vector in the model's input embedding matrix. We define the replacement score as:
\begin{equation}
s_T(\ell,v)=\mathbf{g}_\ell^{\top}\mathbf{w}_v
\label{eq:replacement_score}
\end{equation}
The pair $(\ell,v)$ denotes replacing $t_\ell$ with $v$: $\ell$ specifies the position and $v$ specifies the new token. A larger $s_T(\ell,v)$ indicates that the replacement token is more aligned with the local gradient direction that increases the novelty objective. This score ranks candidate replacements; the actual mean cosine similarity to $E_{\mathrm{seen}}$ is evaluated separately during the periodic stopping check (every $\Delta$ steps).

\begin{algorithm}[!t]
\caption{\qgen with Discrete Token Replacements}
\label{alg:hotflip}
\begin{algorithmic}[1]
\Require Base chunk $\hat{d}_{\mathrm{base}}$, seen embeddings
$E_{\mathrm{seen}}$, threshold $\tau$, maximum query-generation
steps $M$, beam width $B$, replacements per step $S$,
candidate-pool multiplier $k_p$, check interval $\Delta$,
temperature $\mathrm{temp}_s$

\State $T_0 \gets \textsc{Tokenize}(\hat{d}_{\mathrm{base}})$
\State $\mathrm{Beam} \gets \{(T_0,0)\}$
\Comment{Sequence and path score}
\State $\mu_{\mathrm{prev}} \gets +\infty$

\For{step $=1$ to $M$}
    \State $\mathrm{ChildBeam} \gets \emptyset$

    \For{each $(T,\operatorname{PS}(T)) \in \mathrm{Beam}$}
        \State Compute $s_T(\ell,v)$ for all replacements
        $(\ell,v)$
        \State $\mathcal{C}_T \gets$ the top $B \times k_p$
        replacements by $s_T(\ell,v)$
        \State $\boldsymbol{\pi}_T \gets
        \operatorname{softmax}\!\left(
        \left[
        \frac{s_T(\ell,v)}{\mathrm{temp}_s}
        \right]_{(\ell,v)\in\mathcal{C}_T}
        \right)$
        \State $\mathcal{P}_T \gets
        \operatorname{Sample}_B
        \left(\mathcal{C}_T;\boldsymbol{\pi}_T\right)$

        \For{each $(\ell,v)\in\mathcal{P}_T$}
            \State $\mathcal{A} \gets \{(\ell,v)\}$ \Comment{replacement set}
            \State Add to $\mathcal{A}$ up to $S-1$ highest-scoring
            pairs from $\mathcal{C}_T$ at unused positions
            \State $T' \gets \textsc{Replace}(T,\mathcal{A})$
            \State $\operatorname{PS}(T') \gets \operatorname{PS}(T) +
            \displaystyle\sum_{(j,u)\in\mathcal{A}} s_T(j,u)$
            \State $\mathrm{ChildBeam} \gets
            \mathrm{ChildBeam} \cup \{(T',\operatorname{PS}(T'))\}$
        \EndFor
    \EndFor

    \State $\mathrm{Beam} \gets$ the $B$ pairs $(T,\operatorname{PS}(T))$ in
    $\mathrm{ChildBeam}$ with the largest $\operatorname{PS}(T)$

    \If{step mod $\Delta = 0$}
        \State $(T^*,p^*) \gets
        \arg\max_{(T,p)\in\mathrm{Beam}} p$
        \State $\mu \gets
        -\mathcal{L}(T^*)$ \Comment{mean similarity}

        \If{$\mu < \tau$ \textbf{or} $\mu \geq \mu_{\mathrm{prev}}$}
            \State \Return $\textsc{Decode}(T^*)$
        \EndIf

        \State $\mu_{\mathrm{prev}} \gets \mu$
    \EndIf
\EndFor

\State $(T^*,\operatorname{PS}(T^*)) \gets
\arg\max_{(T,\operatorname{PS}(T))\in\mathrm{Beam}} \operatorname{PS}(T)$
\State \Return $\textsc{Decode}(T^*)$
\end{algorithmic}
\end{algorithm}

\subsubsection{Candidate Selection and Beam Update}

A straightforward strategy would be to greedily apply the
highest-scoring replacement to the current text at every step. However, each replacement
score is a local approximation: a replacement that appears optimal at
the current step may lead to a less effective sequence after subsequent
changes. To avoid committing to a single mutation path, \qgen maintains
a beam of $B$ promising sequences.

During beam search, $\operatorname{PS}(T)$ denotes the path score of a token sequence $T$. We initialize $\operatorname{PS}(T_0)=0$, where $T_0$ is the token sequence of $\hat{d}_{\mathrm{base}}$. For each new sequence $T'$ generated from a parent sequence $T$, its
path score is computed by adding the scores of the replacements applied in the current step to $\operatorname{PS}(T)$. The path score is used only to rank beam candidates; the stopping condition is evaluated using the candidate's actual embedding.

For each sequence $T$ in the beam, \qgen ranks all replacements according to $s_T(\ell,v)$ and temporarily retains the top $B \times k_p$ options. Here, $B$ is the beam width and $k_p$ enlarges the set of the next candidates that are sampled. With our default values $B=10$ and $k_p=3$, the algorithm considers the top 30 replacement options for each sequence. This temporary list is rebuilt for every sequence at every step.

From these options, \qgen samples $B$ distinct primary replacements using softmax. Higher-scoring replacements receive higher sampling probabilities, while $\mathrm{temp}_s$ controls how strongly the sampling favors them. Each primary replacement produces one new sequence. When token replacement per step parameter $S>1$, \qgen also applies up to $S-1$ of the highest-scoring remaining replacements at different token positions. We use different positions to prevent the same position from being overwritten more than once during a step.

The algorithm then ranks the resulting sequences by $\operatorname{PS}(T)$ and retains the best $B$ sequences as the next beam. Since the replacement scores accumulated in $\operatorname{PS}(T)$ we can only approximate the effect of each replacement. Algorithm terminates when the mean cosine similarity to $E_{\mathrm{seen}}$ falls below $\tau$, when this similarity stops decreasing, or when the maximum of $M$ steps is reached.

\section{Experimental Evaluation}

\subsection{Experimental Setup}
\label{sec:experimental_setup}

We provide brief explanations of our experimental setup in this section; more details regarding the implementations can be found in Appendix \ref{appendix:exp_setup}. 

\begin{table}[t]
\centering
\caption{Target RAG agent configurations. Agents A and B replicate the setup of~\cite{pirates}. Agent~C is a newly introduced configuration that uses the Neural Bridge RAG dataset. Agents D and E replicate the setup of~\cite{ragcrawler}.}
\label{tab:agents}
\footnotesize
\setlength{\tabcolsep}{1.5pt}
\resizebox{\columnwidth}{!}{%
\begin{tabular}{@{}lllr@{}}
\toprule
\textbf{Agent} & \textbf{LLM} & \textbf{Embedder} & \textbf{KB Dataset} \\
\midrule
A & Llama-3.2-3B-Inst.~\cite{llama} & GTE-large-en~\cite{gte} & Mini-BioASQ \\
B & Phi-3.5-mini-Inst.~\cite{phi} & E5-large-v2~\cite{e5} & Mini-Wikipedia \\
C & Llama-3.2-3B-Inst. & BGE-large-en~\cite{bge} & Neural Bridge RAG \\
D & Llama-3.1-8B-Inst.~\cite{llama} & BGE-large-en & TREC-COVID \\
E & Llama-3.1-8B-Inst. & BGE-large-en & SciDocs \\
\bottomrule
\end{tabular}%
}
\end{table}

\textbf{Target RAG Agents.} To ensure a fair comparison with existing methods, we create an experimental setup similar to PirateRAG~\cite{pirates}. We have a total of 5 agents (A-E); they are summarized in Table~\ref{tab:agents}. 

\textbf{Attacker Configurations.} We use the same common attacker configurations as those in~\cite{pirates}, including the attacker-side embedding model. The attacker uses Snowflake Arctic Embed Large~\cite{snowflake} (\texttt{snowflake-arctic-embed-l}) as its local embedding model. Note that this model is \emph{different} from every target agent's embedder, reinforcing the black-box nature of the attack.

\begin{table}[t]
\centering
\caption{Default values of \qgen parameters. We set these values to balance the attacker's local computational budget with the precision required for exploring the embedding space.}
\label{tab:hotflip_params}
\small
\begin{tabular}{@{}llr@{}}
\toprule
\textbf{Parameter} & \textbf{Symbol} & \textbf{Value} \\
\midrule
Maximum query-generation steps & $M$ & 300 \\
Beam width & $B$ & 10 \\
Token replacements per step & $S$ & 3 \\
Sampling temperature & $\mathrm{temp}_s$  & 0.5 \\
Candidate-pool multiplier & $k_p$ & 3 \\
Check interval & $\Delta$ & 30 \\
Adaptive threshold levels & $n$ & 6 \\
\bottomrule
\end{tabular}
\end{table}

Table~\ref{tab:hotflip_params} lists the default values of \qgen hyperparameters. These values were selected to balance the attacker's local computational budget with the necessary precision of the embedding space exploration. We provide ablation studies on the impact of varying these parameters in Section~\ref{sec:ablation}.

\begin{figure*}[t]
    \centering
    \includegraphics[width=\textwidth]{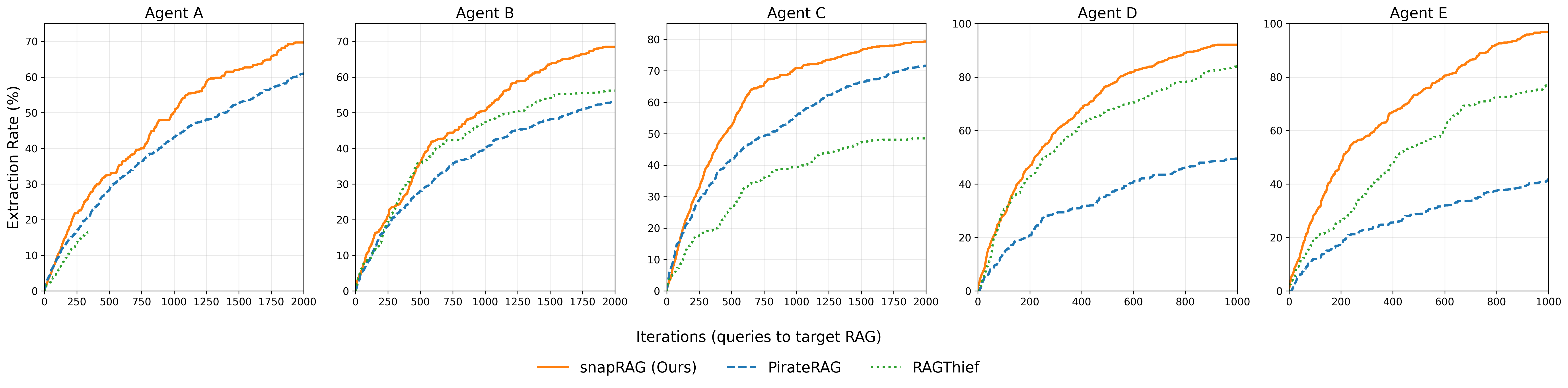}
    \vspace{-16pt}
    \caption{Extraction Rate progression on Agents A-E over the first 2000 and 1000 dispatched queries. \sys (orange line) is above both baselines on every agent configuration. RAGThief auto-terminated on Agent A after only 350 queries.}
    \label{fig:results_abcde}
\end{figure*}

\textbf{Compared Approaches.} We compare our attack against PirateRAG~\cite{pirates} and RAGThief~\cite{ragthief} as they represent the most effective attacks in our domain. Empirical evidence shows that they outperform both DGEA~\cite{dgea} and IKEA~\cite{silentleaks} in terms of the total number of correctly extracted documents in most scenarios. 
In addition, we include RAGCrawler~\cite{ragcrawler} in our comparisons, but we perform the comparison with RAGCrawler in a separate section because, unlike previous approaches that steal target KB word-by-word, RAGCrawler steals semantic information. We detail the reasoning of this decision and the difference in evaluation metrics of RAGCrawler in Section~\ref{section:ragcrawler}.

\textbf{Evaluation Metrics.} We evaluate attacks' extraction performance using the following metrics, consistent with prior work~\cite{pirates, ragthief}.

\underline{Extraction Rate (ER).} The fraction of ground-truth knowledge base chunks $\mathcal{K}$ that are successfully recovered in the $\widehat{\mathcal{K}}$. A chunk $d \in \mathcal{K}$ is considered recovered if the $\widehat{\mathcal{K}}$ contains a chunk $\hat{d}$ such that the ROUGE-L F1 score between $d$ and $\hat{d}$ exceeds a matching threshold $\tau_{\text{match}} = 0.5$~\cite{ragthief, pirates}.

\underline{Query Amount for X\% Extraction (Query@X\%).} Number of API calls required to extract X\% of the target's knowledge base (ER=X\%). A lower Query@X\% indicates a more efficient attack with novel queries that retrieve unexplored regions of the target knowledge base. We typically use X = 50\% in our experiments. 

\underline{Query Efficiency (QE).} To quantify the cost of the attack, we measure the average number of API queries required to extract each previously unseen chunk. Let $|\mathcal{Q}|$ be the total number of adversarial queries dispatched to the target system throughout the attack, and $|\mathcal{K}_{\text{recovered}}|$ be the total number of unique ground truth chunks successfully recovered (ROUGE-L $\ge 0.5$). The Query Efficiency is calculated as:
\begin{equation}
QE = \frac{|\mathcal{Q}|}{|\mathcal{K}_{\text{recovered}}|}
\end{equation}
A lower QE indicates a more cost-efficient attack, demonstrating that the query generator reduces redundant retrievals.

\subsection{Comparison with Competitors}
\label{sec:results}

We evaluate \sys against PirateRAG and RAGThief across all five target agents. Agents~A-C are evaluated under a 2000-query limit, while Agents~D and E are evaluated under a 1000-query limit. Within each agent, all attacks are evaluated using the same maximum number of queries. Figure~\ref{fig:results_abcde} presents the ER progression, Table~\ref{tab:er_results_abcde} reports the final ER and QE results, and Table~\ref{tab:extraction_50_abcde} reports the number of queries required to recover half of each target knowledge base (Query@50\%).

\begin{table*}[!t]
\centering
\caption{Extraction Rate (ER) and Query Efficiency (QE) results for Agents A-E. 
Agents A-C are evaluated under a 2000-query budget, while Agents D and E are evaluated under a 1000-query budget. 
Best results are in bold. Higher ER and lower QE are better.}
\label{tab:er_results_abcde}

\setlength{\tabcolsep}{6pt}
\begin{tabular}{@{} l cc cc cc cc cc @{}}
\toprule

& \multicolumn{2}{c}{\textbf{Agent A}}
& \multicolumn{2}{c}{\textbf{Agent B}}
& \multicolumn{2}{c}{\textbf{Agent C}}
& \multicolumn{2}{c}{\textbf{Agent D}}
& \multicolumn{2}{c}{\textbf{Agent E}} \\
\cmidrule(lr){2-3}
\cmidrule(lr){4-5}
\cmidrule(lr){6-7}
\cmidrule(lr){8-9}
\cmidrule(lr){10-11}

\textbf{Attack}
& ER & QE
& ER & QE
& ER & QE
& ER & QE
& ER & QE \\
\midrule

RAGThief~\cite{ragthief}
& 16.5\%$^{\dagger}$ & 2.12$^{\dagger}$
& 56.2\% & 3.56
& 48.7\% & 4.11
& 84.0\% & 1.19
& 77.0\% & 1.30 \\

PirateRAG~\cite{pirates}
& 61.0\% & 3.28
& 53.2\% & 3.76
& 71.6\% & 2.79
& 51.9\% & 1.93
& 41.7\% & 2.40 \\

\textbf{\sys (Ours)}
& \textbf{69.7\%} & \textbf{2.87}
& \textbf{68.5\%} & \textbf{2.92}
& \textbf{79.3\%} & \textbf{2.52}
& \textbf{92.1\%} & \textbf{1.09}
& \textbf{96.9\%} & \textbf{1.03} \\

\bottomrule
\end{tabular}

\vspace{3pt}
{\raggedright
\scriptsize
$^{\dagger}$ RAGThief auto-terminated on Agent~A after only 350 queries; its ER and QE are reported over that smaller budget.
\par}
\end{table*}

As shown in Figure~\ref{fig:results_abcde}, \sys maintains a clear extraction advantage over both baselines across all five agents. At the end of each evaluation, \sys achieves 69.7\%, 68.5\%, 79.3\%, 92.1\%, and 96.9\% ER on Agents~A-E, respectively (see Table~\ref{tab:er_results_abcde}). Compared with the strongest baseline on each agent, these results correspond to absolute improvements of 8.7, 12.3, 7.7, 8.1, and 19.9 percentage points. Thus, the advantage of \sys ranges from 7.7 to 19.9 percentage points across the evaluated target configurations.

\sys also achieves the lowest QE on every agent among the attacks that complete the full evaluation, as shown in Table~\ref{tab:er_results_abcde}. It requires an average of 2.87, 2.92, and 2.52 queries to recover a new chunk on Agents~A-C and extracts one new chunk per query on Agents~D and E, with QE values of 1.09 and 1.03, respectively. While RAGThief technically recorded a numerically lower QE of 2.12 on Agent A, this is due to the attack auto-terminating after only 350 queries. Early queries are more likely to retrieve undiscovered chunks; therefore, estimating QE under a heavily constrained budget lowers the metric and prevents a fair comparison.

\begin{table}[!t]
\centering
\caption{Number of queries required to extract 50\% of the target knowledge base (Query@50\%). The values in parentheses indicate the relative query cost multiplier compared to \sys. Best results are in bold. Lower is better.}
\label{tab:extraction_50_abcde}

\setlength{\tabcolsep}{3pt}
\renewcommand{\arraystretch}{1.05}

\begin{tabular}{@{}l c c c@{}}
\toprule
\textbf{Agent}
& \shortstack{\textbf{RAGThief}\\\cite{ragthief}}
& \shortstack{\textbf{PirateRAG}\\\cite{pirates}}
& \shortstack{\textbf{\sys}\\\textbf{(Ours)}} \\
\midrule

Agent A
& $\dagger$
& 1379 ($1.38\times$)
& \textbf{997} ($1.00\times$) \\

Agent B
& 1186 ($1.26\times$)
& 1671 ($1.77\times$)
& \textbf{942} ($1.00\times$) \\

Agent C
& 2112 ($4.66\times$)
& 798 ($1.76\times$)
& \textbf{453} ($1.00\times$) \\

Agent D
& 259 ($1.18\times$)
& 923 ($4.21\times$)
& \textbf{219} ($1.00\times$) \\

Agent E
& 411 ($1.98\times$)
& 1409 ($6.77\times$)
& \textbf{208} ($1.00\times$) \\

\bottomrule
\multicolumn{4}{@{}l}{%
\footnotesize\textit{$\dagger$: The attack automatically terminated before reaching 50\% extraction.}}
\end{tabular}
\end{table}

Finally, \sys demonstrates significant cost efficiency when extracting half of the target databases. Since commercial RAG APIs typically bill by usage, minimizing the total number of dispatched queries implies a cheaper attack. As detailed in Table~\ref{tab:extraction_50_abcde}, our attack recovers half of the target knowledge base using 997, 942, 453, 219, and 208 queries on Agents~A-E, respectively. It is worth noting that since each target knowledge base contains exactly 1000 documents, our attack successfully extracts half of the data with fewer than 1000 API interactions. The strongest baseline on each agent requires 1.38$\times$, 1.26$\times$, 1.76$\times$, 1.18$\times$, and 1.98$\times$ as many queries to reach the same extraction level. RAGThief does not reach 50\% extraction on Agent~A before automatically terminating. Overall, \sys is the first attack to reach 50\% extraction on every agent. We detail the estimated cost comparison based on API prices per token in Appendix~\ref{appendix:cost}.

Together, these results show that the extraction and query-efficiency advantages of \sys persist across target agents with different knowledge bases, LLMs, embedding models, and top-$k$ retrieval configurations.

\subsection{Evaluation Against Active Defense}
\label{sec:active_defense}

\textbf{Defense setup.} To evaluate the attacks against a target-side active defense, we augment each RAG agent with Llama Guard~3 8B~\cite{llama,inan2023llama}. After retrieval, the target constructs the full RAG prompt containing the retrieved documents and the user's query, including any appended injection command. The guard first moderates this full prompt before target LLM generation. If the input is classified as unsafe, the target LLM is not executed, and the endpoint returns a fixed block response. For inputs classified as safe, the target LLM generates a response. Then, the guard moderates this response, including the same retrieved documents. If it classifies it as unsafe, the output is replaced by the same block response. Every call, including a blocked call, consumes one query from the attack budget and yields no extracted chunks. We evaluate \sys, PirateRAG, and RAGThief against these actively defended RAG agents using the same query budgets as in the previous experiments. 

\begin{figure*}[t]
    \centering
    \includegraphics[width=\textwidth]{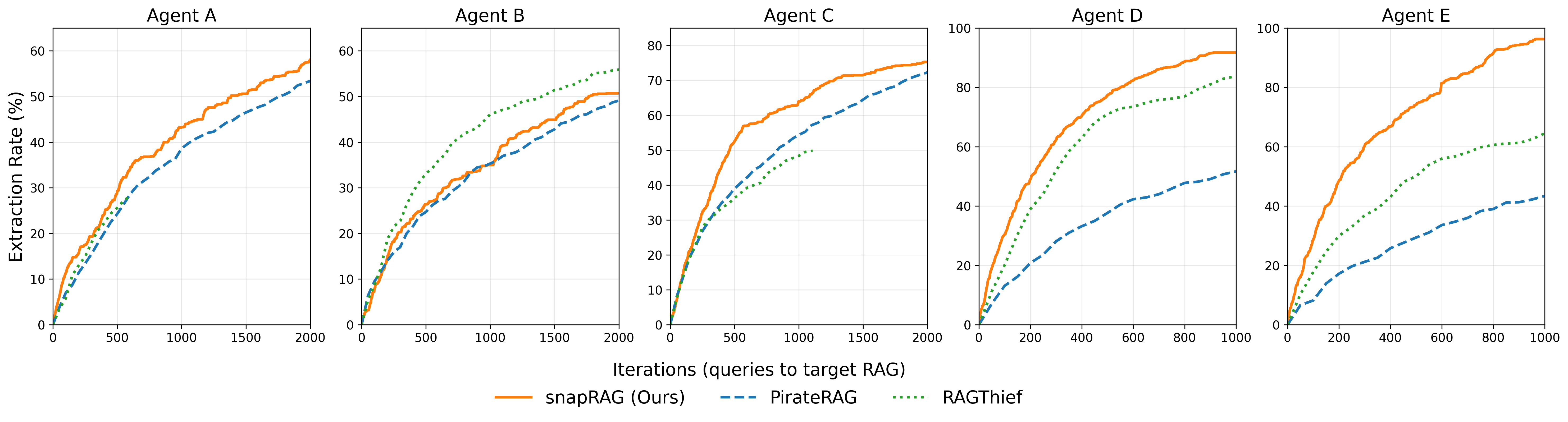}
    \vspace{-16pt}
    \caption{Extraction Rate progression for Agents A-E under defense. RAGThief exhausts its query queues on Agents A and C after 591 and 1107 queries, respectively.}
    \label{fig:guarded_results}
\end{figure*}

\textbf{Results and discussion.} Figure \ref{fig:guarded_results} reports the ER progression versus the number of queries to the defended RAG agents. \sys reaches 58.0\%, 50.7\%, 75.3\%, 91.8\%, and 96.3\% ER on Agents A-E, respectively, leading both competitors on four of the five agents. Agent B is the only exception: RAGThief obtains 55.9\% ER, while \sys reaches 50.7\% and PirateRAG reaches 49.1\%. Thus, \sys exceeds PirateRAG on every defended agent and remains second on the only agent where it does not lead.

\begin{table}[t]
\centering
\small
\setlength{\tabcolsep}{3pt}
\caption{\sys rejection rates under defense. Percentages use the query budget as the denominator.}
\label{tab:guarded_blocks}
\resizebox{\columnwidth}{!}{%
\begin{tabular}{@{}lrrrr@{}}
\toprule
\textbf{Agent} & \textbf{Budget} & \textbf{Input} & \textbf{Output} & \textbf{Total} \\
\midrule
Agent A& 2000 & 323 (16.15\%) & 10 (0.50\%) & 333 (16.65\%) \\
Agent B & 2000 & 684 (34.20\%) & 132 (6.60\%) & 816 (40.80\%) \\
Agent C & 2000 & 195 (9.75\%) & 21 (1.05\%) & 216 (10.80\%) \\
Agent D & 1000 & 57 (5.70\%) & 10 (1.00\%) & 67 (6.70\%) \\
Agent E & 1000 & 11 (1.10\%) & 1 (0.10\%) & 12 (1.20\%) \\
\bottomrule
\end{tabular}%
}
\end{table}

Table~\ref{tab:guarded_blocks} shows that the fraction of blocked \sys queries differs substantially across agents, ranging from 1.20\% on Agent E to 40.80\% on Agent B. Input moderation accounts for most blocks on every agent, although output moderation blocks 6.60\% of all queries on Agent B. Even with 40.80\% of its queries blocked, \sys recovers 50.7\% of Agent B's knowledge base. These results show that an actively defended RAG agent raises the attack cost and reduces ER in some settings, but does not consistently prevent knowledge base recovery. We further analyze the guard's rejection classes and their variation across agents in Appendix~\ref{appendix:guard_analysis}.

We note that for \sys, the ER changes by $-11.7$, $-17.8$, $-4.0$, $-0.3$, and $-0.6$ percentage points on Agents A-E under the defense, respectively, compared to the no-defense scenario. The largest difference occurs on Agent B, whereas the outcomes for Agents D and E remain close to their no-defense results, which shows a consistent relation with their block rates (see Table~\ref{tab:guarded_blocks}).

\subsection{Comparison with RAGCrawler}
\label{section:ragcrawler}

\textbf{Evaluation setup.}
As discussed in Section~\ref{sec:experimental_setup}, we evaluate
RAGCrawler~\cite{ragcrawler}, the most recent work, separately because its objective differs
from that of \sys, PirateRAG, and RAGThief. The latter attacks attempt
to reconstruct the target knowledge base at the chunk level by commanding
the target LLM to reproduce its retrieved context. RAGCrawler instead
performs semantic knowledge stealing: it systematically explores the
target knowledge base and extracts information exposed through
natural language answers, without requiring the retrieved chunks to be
reproduced word for word. Consequently, Extraction Rate (ER), which
requires a stolen chunk to attain a ROUGE-L score of at least 0.5
against a ground-truth chunk, does not fully capture RAGCrawler's
intended objective. We nevertheless report its ER for completeness in
Table~\ref{tab:ragcrawler_er_cr}.

Following RAGCrawler, we therefore also evaluate the attacks using
Coverage Rate (CR). For a query $Q$, let
$\mathcal{K}_{\mathrm{ret}}(Q) \subseteq \mathcal{K}$ denote the set
of ground-truth knowledge-base chunks internally retrieved by the
target RAG. We calculate CR as
\begin{equation}
\mathrm{CR}
=
\frac{
\left|
\displaystyle\bigcup_{Q \in \mathcal{Q}_{\mathrm{useful}}}
\mathcal{K}_{\mathrm{ret}}(Q)
\right|
}{
|\mathcal{K}|
},
\end{equation}
where $\mathcal{Q}_{\mathrm{useful}}$ contains the queries that yield a
useful target response. This calculation is performed by the
offline evaluator using the target's ground-truth retrieval logs; these
retrieved chunks are not exposed to the attacker during the attack.

For \sys, PirateRAG, and RAGThief, a response is considered useful when
the corresponding extraction mechanism parses at least one chunk from
the target LLM's output. Thus, retrieved documents associated with
refusals, natural answers without leaked chunks, or responses that
cannot be parsed are excluded from CR. For RAGCrawler, we follow its
native evaluation methodology and use its original implementation:
retrievals are counted only when the algorithm marks the
response as valid for coverage. In all cases, duplicate retrieved
documents are counted only once, and the denominator ($|\mathcal{K}|$) is the 1,000
documents in the target knowledge base. Unlike ER, CR measures how much
of the target knowledge base is reached by useful attack queries,
irrespective of whether each retrieved document is reproduced
verbatim.

\begin{table*}[!t]
\centering
\caption{Extraction Rate (ER) and Coverage Rate (CR) for Agents A-E
under a 1000-query budget. Higher is better. Best results for each
agent and metric are in bold.}
\label{tab:ragcrawler_er_cr}

\setlength{\tabcolsep}{4.5pt}
\renewcommand{\arraystretch}{1.05}
\small

\begin{tabular}{@{}l cc cc cc cc cc@{}}
\toprule
& \multicolumn{2}{c}{\textbf{Agent A}}
& \multicolumn{2}{c}{\textbf{Agent B}}
& \multicolumn{2}{c}{\textbf{Agent C}}
& \multicolumn{2}{c}{\textbf{Agent D}}
& \multicolumn{2}{c}{\textbf{Agent E}} \\
\cmidrule(lr){2-3}
\cmidrule(lr){4-5}
\cmidrule(lr){6-7}
\cmidrule(lr){8-9}
\cmidrule(lr){10-11}

\textbf{Attack}
& ER & CR
& ER & CR
& ER & CR
& ER & CR
& ER & CR \\
\midrule

RAGThief~\cite{ragthief}
& 16.5\%$^{\dagger}$ & 18.0\%$^{\dagger}$
& 47.4\% & 49.4\%
& 39.4\% & 40.7\%
& 84.0\% & 86.2\%
& 77.0\% & 78.7\% \\

PirateRAG~\cite{pirates}
& 43.1\% & 46.0\%
& 40.1\% & 40.6\%
& 55.7\% & 55.8\%
& 51.9\% & 57.8\%
& 41.7\% & 45.4\% \\

RAGCrawler~\cite{ragcrawler}
& 0.1\% & 49.0\%
& 1.0\% & \textbf{63.9\%}
& 0.4\% & 38.4\%
& 0.2\% & 73.4\%
& 0.5\% & 84.1\% \\

\textbf{\sys{} (Ours)}
& \textbf{50.2\%} & \textbf{50.6\%}
& \textbf{50.6\%} & 51.5\%
& \textbf{70.8\%} & \textbf{71.9\%}
& \textbf{92.1\%} & \textbf{92.6\%}
& \textbf{96.9\%} & \textbf{97.3\%} \\

\bottomrule
\end{tabular}

\vspace{3pt}
{\raggedright
\scriptsize
$^{\dagger}$ RAGThief automatically terminated on Agent~A after
350 queries; its ER and CR are therefore reported at its terminal
query budget. All other results are reported at 1,000 queries.
\par}
\end{table*}

\textbf{Results and discussion.}
Table~\ref{tab:ragcrawler_er_cr} shows that \sys obtains the highest ER in all,
and highest CR in four out of five agent configurations.
Its CR values are 50.6\%, 51.5\%, 71.9\%, 92.6\%, and 97.3\% on
Agents~A-E, respectively. Compared with the strongest available
competitors, these results represent CR
improvements of 1.6, 16.1, 6.4, and 13.2 percentage points for agents A, C, D, E. We note that \sys is second strongest attack in agent B, where RAGCrawler achieves 63.9\% CR.

We observe that RAGCrawler achieves at most 1.0\% ER across all five agents, despite attaining
substantial CR values ranging from 38.4\% to 84.1\%. This difference follows directly from its semantic extraction objective. RAGCrawler's queries reach a large fraction of
the target knowledge base, but the target LLM generally synthesizes
the retrieved information into natural answers rather than reproducing
individual source chunks. These answers can contain knowledge derived
from several retrieved documents while having little word-level
overlap with any complete ground truth chunk. They are therefore
credited by RAGCrawler's CR metric but rejected by the stricter
ROUGE-L based ER metric. Accordingly, the low ER values should not be
interpreted as an inability to explore the target knowledge base;
instead, they demonstrate that ER is unsuitable as the sole measure of
a semantic knowledge stealing attack.

The ER and CR values of \sys, PirateRAG, and RAGThief are much closer
because these attacks explicitly prompt the target LLM to reproduce
retrieved chunks. For \sys, the difference between CR and ER remains
below 1.1 percentage point on every agent. This indicates that nearly
every distinct document reached by a useful \sys query is also
successfully reconstructed. Together,
these results show that \sys combines broad knowledge-base exploration
with reliable word-level recovery, achieving both the highest coverage
and the highest extraction rate.

\subsection{Ablation Studies}
\label{sec:ablation}

We perform ablation studies by varying one parameter at a time while keeping others fixed at their default values from Table~\ref{tab:hotflip_params}. All ablation experiments are run on Agent C. We report results using three primary metrics: ER under a 1000-query budget, the number of queries needed to extract half the database (Query@50\%), and median generation times. For time, we measure both the seconds required to generate a single query (Query(s)) and to generate $n$ queries from a single base chunk (Base(s)). Table~\ref{tab:ablation_budget} shows the ER and Query@50\% results for each parameter, and Table \ref{tab:generation_time} shows the Base(s) and Query(s) results. We also present additional runtime measurements of our attack in Appendix~\ref{appendix:cost}.

\begin{table}[!t]
\centering
\caption{Ablation on \sys parameters. The default value of each parameter is in bold. ER is measured with 1000 queries (or at the last iteration if the method terminated earlier). Lower Query@50\% and higher ER are better.}
\label{tab:ablation_budget}
\begin{tabular}{@{}cccc@{}}
\toprule
\textbf{Parameter} & \textbf{Value} & \textbf{ER} & \textbf{Query@50\%} \\
\midrule
\multirow{3}{*}{$M$}
 & 200         & 67.7\% & 410 \\
 & \textbf{300}    & 70.8\% & 453 \\
 & 400             & 68.2\% & 430 \\
\midrule
\multirow{3}{*}{$B$}
 & 6               & 68.9\% & 412 \\
 & \textbf{10}     & 70.8\% & 453 \\
 & 14              & 66.1\% & 453 \\
\midrule
\multirow{3}{*}{$S$}
 & 1               & 69.1\% & 410 \\
 & \textbf{3}      & 70.8\% & 453 \\
 & 5               & 70.2\% & 430 \\
\midrule
\multirow{3}{*}{$n$}
 & 4               & 64.1\% & 461 \\
 & \textbf{6}      & 70.8\% & 453 \\
 & 8               & 66.8\% & 423 \\
\midrule
\multirow{3}{*}{$\tau_{\mathrm{rouge}}$}
 & \textbf{0.55}   & 70.8\% & 453 \\
 & 0.65            & 66.9\% & 449 \\
 & 0.75            & 69.0\% & 447 \\
\bottomrule
\end{tabular}
\end{table}

\textbf{Maximum steps $M$.} The \qgen algorithm exhibits strong robustness with respect to the maximum step parameter $M$, with only a minimal effect on overall extraction performance. As seen in Table~\ref{tab:ablation_budget}, ER fluctuates by only 2.6–3.1\%, and Query@50\% remains stable between 410 and 453. More importantly, query generation times remain largely consistent across varying $M$ values (see Table~\ref{tab:generation_time}). This indicates that the algorithm mostly terminates via the early stopping mechanism, which is triggered upon reaching the target dissimilarity threshold $\tau$ (see Algorithm~\ref{alg:hotflip}), rather than exhausting the step budget. Consequently, an attacker can safely initialize $M$ to a high value to guarantee sufficient exploratory capacity; the early stopping condition naturally prevents any unnecessary processing overhead.

\textbf{Beam width $B$}.
Similarly, extraction performance is largely insensitive to variations in $B$. While $B$ defines the width of the search space, where smaller values approximate a greedy search, modifying it yields no substantial changes in ER or Query@50\%. However, $B$ has a direct and observable effect on generation times. From Table~\ref{tab:generation_time}, we observe that query generation requires only 28.84 s at $B=6$, but this latency more than doubles to 68.83 s at $B=14$. Therefore, to optimize computational efficiency without compromising extraction capabilities, selecting a lower $B$ is preferable.

\textbf{Token replacements per step $S$}. The parameter $S$ sets how many modifications the \qgen algorithm applies to the base chunk in a single step. Similar to $B$, changing $S$ does not show a clear effect in Table~\ref{tab:ablation_budget}. However, increasing $S$ causes a large increase in computational time. Specifically, the per-query generation time jumps from 8.60 s at $S=1$ to 50.39 s at $S=5$ (Table~\ref{tab:generation_time}). This shows that lower values of $S$ are better for keeping the execution fast.

Overall, the results for the \qgen parameters $M$, $B$, and $S$ demonstrate that while the algorithm's extraction capabilities are robust to minor parameter changes, its computational overhead is sensitive to them. Based on these findings, an attacker should configure these parameters according to their constraints. If the objective is to maximize the ER, our evaluations indicate that moderate parameter values (e.g., $M=300$, $B=10$, $S=3$) offer a stable performance.

\begin{table}[!t]
\centering
\caption{Generation times (in seconds) per single query (Query) and multiple queries from a base chunk (Base) across different parameter configurations.}
\label{tab:generation_time}
\setlength{\tabcolsep}{2pt}
\begin{tabular}{@{}ccccc@{}}
\toprule
\textbf{Parameter} & \textbf{Value} & \textbf{Base(s)} & \textbf{Query(s)} \\
\midrule
\multirow{2}{*}{$M$}
 & 200     & 138.10 & 23.02\\
 & 400       & 137.85 & 22.97\\
\midrule
\multirow{2}{*}{$B$}
 & 6      & 173.04 & 28.84\\
 & 14     & 412.99 & 68.83 \\
\midrule
\multirow{2}{*}{$S$}
 & 1        & 51.60 & 8.60\\
 & 5       & 302.32 & 50.39\\
\midrule
\multirow{2}{*}{$n$}
 & 4    & 68.90 & 17.23 \\
 & 8    & 135.51 & 16.94 \\
\bottomrule
\end{tabular}
\end{table}

\textbf{Adaptive threshold level $n$} affects the number of modified query variants generated from each extracted chunk. A higher $n$ will result in more API calls to the target system and thus increase the cost. As shown in Table~\ref{tab:ablation_budget}, higher $n$ yields a slight improvement in Query@50\% results, modestly reducing the number of queries needed to reach the 50\% extraction. The overall ER, however, does not exhibit a clear correlation with $n$, demonstrating only minor fluctuations of 4\% to 6.7\%. However, $n$ has a definitive impact on generation times. From Table~\ref{tab:generation_time}, we observe that although the generation time per query remains similar, the Base(s) value almost doubles from 68.90 s to 135.51 s as $n$ increases from 4 to 8. This demonstrates that the total computation time required to process a single extracted base chunk increases substantially as more query variants are requested. 

\textbf{ROUGE-L deduplication threshold $\tau_{\mathrm{rouge}}$} controls how aggressively the Stolen KB rejects near-paraphrases of already-stored chunks. A higher $\tau_{\mathrm{rouge}}$ lets more chunks slip through, which might cause redundant query generation and waste attacker budget, while a lower threshold might reject chunks that are actually new. Our ablation results show no significant difference in performance across the tested $\tau_{\mathrm{rouge}}$ values of 0.55, 0.65, and 0.75. Query@50\% ranges narrowly from 447 to 453, showing that the extraction pipeline is robust to $\tau_{\mathrm{rouge}}$ within a standard operational range, allowing an attacker to rely on a moderate default value (e.g., $\tau_{\mathrm{rouge}} = 0.55$). 

\begin{figure}[t]
    \centering
    \includegraphics[width=0.8\linewidth]{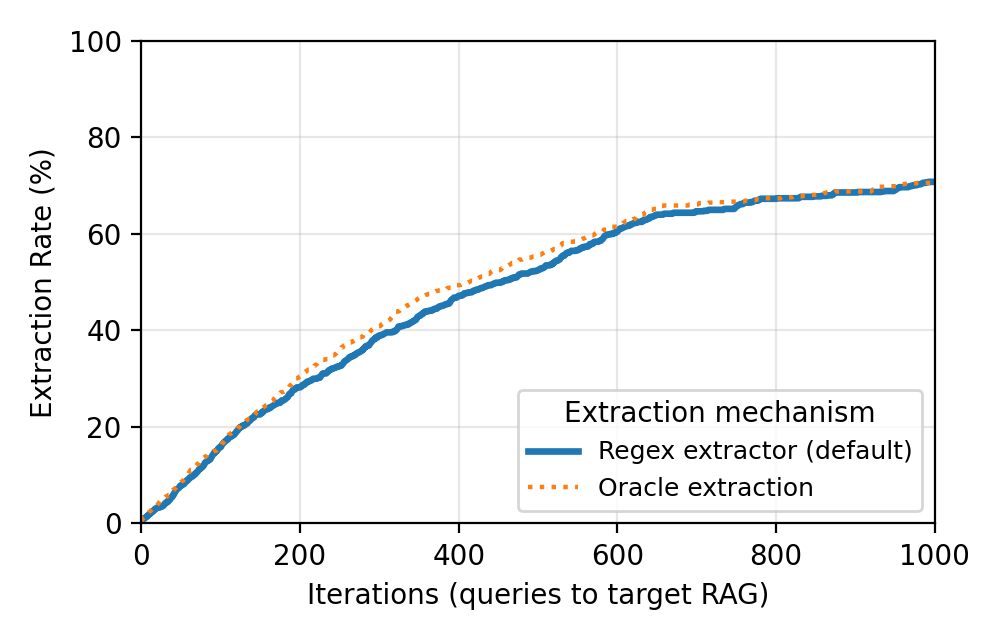}
    \vspace{-8pt}
    \caption{Default regex-based extractor versus an oracle that returns retrieved chunks without parsing error. The two curves are nearly indistinguishable, which shows that our extraction mechanism with our injection command successfully steals chunks from the LLM's response.}
    \label{fig:ablation_extraction}
\end{figure}

\textbf{Extraction mechanism.} To demonstrate the effectiveness of our extraction mechanism when combined with the injection command, we compare it against a hypothetical oracle baseline. This oracle always returns the chunks that the target LLM uses to generate responses. This eliminates any extraction failures and LLM hallucinations. Figure~\ref{fig:ablation_extraction} demonstrates that both settings follow a similar performance, which proves our extraction mechanism's and injection command's effectiveness. To extract half of the target database, Oracle extraction achieves 412 Query@50\%, and the regex extractor achieves 453 Query@50\%, which presents only a slight improvement due to the absence of extraction errors. We present the results of the compared attacks combined with our extraction mechanism in Appendix~\ref{appendix:qgen_eval}, showing that it improves performance in some scenarios.

\section{Conclusion}

In this work, we introduced \sys: a novel, black-box knowledge extraction attack on RAG systems that eliminates the dependency on an external LLM. Unlike prior methods that rely on external LLMs for query generation, which naturally require extra resources, our method requires significantly fewer queries to extract knowledge from RAG systems. By combining a discrete token mutation strategy with an adaptive dissimilarity threshold, a multi-level deduplication filter, and a ranked dispatch mechanism, our method systematically generates queries in unexplored semantic regions of the target knowledge base and thereby ensures the attacker's budget is spent effectively. Experiments show that \sys consistently outperforms state-of-the-art LLM-dependent attacks under limited query budgets, achieving higher ER and QE. Under an active input and output moderation defense, \sys remains the leading attack on four of five targets and recovers 50.7-96.3\% of their knowledge bases. 

Although moderation reduces extraction in some settings, it does not consistently prevent recovery. This finding exposes a substantial privacy risk in current RAG systems and motivates defenses designed specifically to prevent such adaptive extraction attacks. We plan to work on defenses in future work. We also plan to extend \sys to graph-based \cite{Edge2024FromLT, hu2025grag, sen-etal-2023-knowledge} and multi-modal RAG systems, where different retrieval mechanisms and prompt structures may introduce new challenges for knowledge extraction.

\bibliographystyle{plainurl}
\bibliography{references}

\appendix

\section{Ethical Considerations}

In this paper, we present a practical knowledge extraction attack on RAG systems. Although \sys is an attack method, we believe that publicly disclosing these vulnerabilities is crucial to advancing AI safety. By demonstrating how efficiently data can be extracted, we aim to highlight the practical security vulnerabilities of current RAG systems, and we hope that our research will encourage the development of stronger defenses in future RAG pipelines.

To ensure that our research was conducted ethically, all experiments were run in safe, locally hosted environments. We did not target commercial RAG systems, live APIs, or private user data. Additionally, the databases we attacked were built entirely from public datasets, generating a simulation environment. This allowed us to demonstrate the effectiveness of our attack without putting real sensitive information at risk.

\section{More Details Regarding Experiment Setup} \label{appendix:exp_setup}

\subsection{Target RAG Agents} 

To ensure a fair comparison with existing methods, we create an experimental setup similar to PirateRAG~\cite{pirates}. We define various target RAG agents, each of which is a combination of various LLMs, embedding models, the number of retrieved documents (top-k), and private databases with different domains. Table~\ref{tab:agents} summarizes our configurations. All agents follow the standard retrieve-then-generate pipeline: given a user query, the system retrieves the top-$k$ most relevant documents from its knowledge base using the specified embedder, injects them into a prompt template, and generates a response with the LLM. We reproduce two agents of PirateRAG, two agents from the experimental setup of RAGCrawler, and add one additional agent with a new dataset to evaluate generalization. All agents are served via vLLM~\cite{vllm} for efficient inference, exposing a single HTTP endpoint that accepts a query and returns the generated response.

\textbf{Agent~A (BioASQ)} uses 1,000 biomedical passages from the Mini-BioASQ,\footnote{\url{https://huggingface.co/datasets/rag-datasets/rag-mini-bioasq}} embedded with Alibaba-NLP/gte-large-en-v1.5~\cite{gte}, retrieving $k =3$ documents per query. The LLM is Meta Llama-3.2-3B-Instruct~\cite{llama}.

\textbf{Agent~B (Wikipedia)} uses 1,000 passages from the Mini-Wikipedia,\footnote{\url{https://huggingface.co/datasets/rag-datasets/rag-mini-wikipedia}} embedded with intfloat/e5-large-v2~\cite{e5}, with $k = 5$ retrievals. The LLM is Microsoft Phi-3.5-mini-instruct~\cite{phi}.

\textbf{Agent~C (Neural Bridge)} is a new agent we introduce to evaluate generalization beyond the datasets used in prior work. Following the creation mechanism of other agents~\cite{pirates}, we construct a knowledge base of 1,000 passages sampled from the Neural Bridge RAG dataset,\footnote{\url{https://huggingface.co/datasets/neural-bridge/rag-dataset-12000}} a diverse collection of question--context--answer triples spanning multiple domains.
The embedding model is BAAI/bge-large-en-v1.5~\cite{bge} with $k =5$ retrievals, and the LLM is Meta Llama-3.2-3B-Instruct.

\textbf{Agents~D and E (TREC-COVID and SciDocs)} use 1,000 passages from the TREC-COVID and SciDocs datasets, which are from the BEIR~\cite{beir} benchmark. Following the experimental setup of~\cite{ragcrawler}, these settings use the BGE family embedding model, Meta Llama-3.1-8B-Instruct~\cite{llama} LLM, with $k = 10$.

\subsection{Attacker Configuration} 

To ensure a fair comparison, we used the same common attacker configurations as those in~\cite{pirates}, including the attacker-side embedding model and deduplication parameters. The attacker uses Snowflake Arctic Embed Large~\cite{snowflake} (\texttt{snowflake-arctic-embed-l}) as its local embedding model for all operations: Stolen~KB indexing, query history tracking, and computations in \qgen. This model is \emph{different} from every target agent's embedder, reinforcing the black-box nature of the attack. We set the semantic deduplication threshold to $\tau_{\text{dedup}} = 0.95$. The ROUGE-L deduplication threshold is set to $\tau_{\text{rouge}} = 0.55$, checked against the $k_r{=}5$ nearest neighbours. The query-level deduplication threshold is $\tau_{\text{qdedup}} = 0.95$.

Table~\ref{tab:hotflip_params} provides the default values of the hyperparameters for our \qgen algorithm. We select these values to balance the attacker's local computational budget with the necessary precision of the embedding space exploration. Parameters $M$ and $B$ determine the granularity of the search, controlling how finely the algorithm navigates toward the novelty objective. The number of adaptive threshold levels $n$ is determined by the attacker's financial constraints regarding the target API. Since $n$ sets a hard upper bound on the number of novel queries generated per extracted chunk, it allows the attacker to explicitly control their query spending in relation to the amount of stolen data.

\begin{table*}[t]
\centering
\small
\setlength{\tabcolsep}{5pt}
\caption{Llama Guard~3 hazard classes assigned to \sys queries. Counts include unsafe input and output decisions. Guard execution errors are excluded from the class rows and presented separately.}
\label{tab:guard_categories}
\begin{tabular}{@{}clrrrrr@{}}
\toprule
\textbf{Class} & \textbf{Hazard category} & \textbf{Agent A} & \textbf{Agent B} & \textbf{Agent C} & \textbf{Agent D} & \textbf{Agent E} \\
\midrule
S1  & Violent Crimes             & 1   & 16  & 8  & 1  & 1 \\
S2  & Non-Violent Crimes         & 1   & 1   & 11 & 0  & 0 \\
S3  & Sex-Related Crimes         & 0   & 0   & 0  & 0  & 0 \\
S4  & Child Sexual Exploitation  & 7   & 16  & 13 & 0  & 0 \\
S5  & Defamation                 & 3   & 120 & 30 & 0  & 0 \\
S6  & Specialized Advice         & 248 & 4   & 26 & 61 & 1 \\
S7  & Privacy                    & 3   & 23  & 49 & 0  & 1 \\
S8  & Intellectual Property      & 5   & 142 & 41 & 1  & 5 \\
S9  & Indiscriminate Weapons     & 16  & 6   & 1  & 3  & 1 \\
S10 & Hate                       & 6   & 405 & 24 & 1  & 2 \\
S11 & Suicide \& Self-Harm       & 2   & 4   & 1  & 0  & 0 \\
S12 & Sexual Content             & 1   & 5   & 2  & 0  & 0 \\
S13 & Elections                  & 3   & 44  & 9  & 0  & 1 \\
S14 & Code Interpreter Abuse     & 15  & 30  & 1  & 0  & 0 \\
\midrule
\textbf{--} & \textbf{Unsafe classifications} & \textbf{311} & \textbf{816} & \textbf{216} & \textbf{67} & \textbf{12} \\
\textbf{--} & \textbf{Guard errors}           & \textbf{22}  & \textbf{0}   & \textbf{0}   & \textbf{0}  & \textbf{0} \\
\textbf{--} & \textbf{Total blocked calls}    & \textbf{333} & \textbf{816} & \textbf{216} & \textbf{67} & \textbf{12} \\
\bottomrule
\end{tabular}
\end{table*}

\subsection{Compared Approaches}
\label{sec:baselines}

For experimental comparisons, we use the open-source implementations provided by~\cite{pirates} for RAGThief and PirateRAG. For RAGThief~\cite{ragthief}, we use the provided implementation directly. For PirateRAG~\cite{pirates}, we restrict the algorithm to use only the first injection command from its default injection pool. This adjustment is necessary to ensure a fair comparison in budget-limited scenarios; evaluating a pool of multiple injection commands for a single query inherently multiplies the number of API calls made to the target, and doing so severely inflates the query budget for PirateRAG. For RAGCrawler~\cite{ragcrawler}, we design Agents D and E to replicate its two experimental setups. We use the official implementation of \cite{ragcrawler}, with Qwen2.5-7B-Instruct~\cite{qwen2} as the attacker side LLM, and discuss the results separately from the other works in Section~\ref{section:ragcrawler} because of the stated reasons.

\section{Analysis of Guard Rejections}
\label{appendix:guard_analysis}

To investigate the reasons behind guard rejections and understand agent-to-agent different in guard rejection rates, we study the moderation metadata. Llama Guard 3 assigns unsafe inputs and outputs to 14 hazard classes~\cite{llama}. Table~\ref{tab:guard_categories} presents these classifications across Agents A-E. Each unsafe decision in our runs contains exactly one class, so the class counts sum to the number of unsafe classifications. These classes are model-generated moderation decisions rather than human-validated labels of the dispatched query or retrieved content. The 22 guard execution errors on Agent A are reported separately.

The biomedical configurations show concentrated class distributions. On Agent A, S6 (Specialized Advice) accounts for 248 of 311 unsafe classifications (79.7\%). On Agent D, it accounts for 61 of 67 (91.0\%). This concentration is consistent with the medical and scientific passages retrieved by these agents being evaluated as part of the full prompt. However, Agent A and Agent D have different total block rates (16.65\% and 6.70\%, respectively). A shared broad domain is therefore not sufficient to predict rejection; the exact combination of the retrieved passages, query, injection command, and, for output moderation, generated response may also contribute.

Agent B has the highest rejection rate and a high-volume distribution across several classes. S10 (Hate) accounts for 405 of its 816 unsafe classifications (49.6\%), followed by S8 (Intellectual Property) with 142 (17.4\%) and S5 (Defamation) with 120 (14.7\%). Furthermore, 124 of its 132 output blocks are assigned S10. One plausible explanation is the context of the Wikipedia-based knowledge base: retrieved passages cover people, historical conflicts, protected groups, elections, and other topics that interact with several guard classes. The fixed injection command also requests verbatim reproduction, which contributes to S8 decisions. However, the same command is used for every agent, so it cannot by itself explain Agent B's higher rejection rate. These observations support the impact of retrieved context in the final guard decision.

Agent C also reflects its mixed-domain configuration: its most frequent classes are S7 (Privacy, 49), S8 (Intellectual Property, 41), S5 (Defamation, 30), S6 (Specialized Advice, 26), and S10 (Hate, 24), without a single class dominating as strongly as on Agents A, B, or D. Agent E produces only 12 unsafe classifications, of which five are S8. This count is too small to support a strong conclusion about its class distribution. 

Overall, the results suggest that the guard's rejection patterns are knowledge base-dependent, with different safety categories occurring more frequently across different domains, influenced by the nature of the underlying knowledge base.

\section{Evaluation of Query Generation Strategies} \label{appendix:qgen_eval}

\textbf{Evaluation setup.} To isolate the performance of the underlying query generation algorithms from the effectiveness of prompt injection payloads, we introduce two modified attack variants: PirateRAG+\textsc{snap} and RAGThief+\textsc{snap}. In these variants, we replace their default injection commands and extractors with our own, indicated as "+\textsc{snap}". This decoupling ensures that the empirical comparison is entirely independent of the design of the injection commands, enabling a direct evaluation of each method's query generation strategy.

\begin{table}[t]
\centering
\caption{Number of queries required to extract 50\% of the target knowledge base (Query@50\%). The values in parentheses indicate the relative query cost multiplier compared to \sys. Best results are in bold. Lower is better.}
\label{tab:extraction_50_appendix}
\resizebox{\columnwidth}{!}{%
\begin{tabular}{l l l l}
\toprule
\textbf{Attack} & \textbf{Agent A} & \textbf{Agent B} & \textbf{Agent C} \\
\midrule
RAGThief+\textsc{snap} & 3409 ($3.42\times$)         & $\dagger$                   & 2624 ($5.79\times$) \\
PirateRAG+\textsc{snap} & 1291 ($1.29\times$) & \textbf{855} ($0.91\times$) & $\ddagger$ \\
\textbf{\sys (Ours)} & \textbf{997} ($1.00\times$) & 942 ($1.00\times$) & \textbf{453} ($1.00\times$) \\
\bottomrule
\multicolumn{4}{l}{\small \textit{$\dagger$ : The attack automatically terminated before reaching 50\% extraction.}} \\
\multicolumn{4}{l}{\small \textit{$\ddagger$ : We halted the attack after 10,000 queries as it failed to reach 50\% extraction.}} \\
\end{tabular}%
}
\end{table}

\textbf{Results and discussion.} Table~\ref{tab:extraction_50_appendix} demonstrates that although incorporating our extraction mechanism slightly degrades performance in some cases, it substantially improves PirateRAG’s results on Agents A and B. Specifically, PirateRAG+\textsc{snap} reaches 50\% extraction with only 1291 and 855 queries to the target RAG, compared to the original PirateRAG's 1379 and 1671 queries on Agents A and B, respectively. We observe that while \sys still outperforms the +\textsc{snap} versions of its competitors on Agents A and C, PirateRAG+\textsc{snap} performs better on Agent B, requiring 0.91$\times$ fewer queries than \sys. On the other hand, in Table \ref{tab:abc_agents_appendix}, we observe that under a 2000 query budget, \sys outperforms all competitors with 69.7\%, 68.5\%, and 79.3\% ER, compared to the closest results of 57.4\%, 68.1\%, and 49.2\% on Agents A, B, and C, respectively. In this table, we can also see that the results for RAGThief+\textsc{snap} significantly increase from 16.5\% to 44.6\% compared to RAGThief, and the attack does not terminate early on Agent A.

These results underscore the substantial standalone value of our query generation and extraction mechanisms. The fact that PirateRAG’s query cost is almost cut in half by adopting our prompt injection payload shows that our extraction framework is highly robust and transferable. This proves that \sys offers two distinct, valuable contributions: a budget-efficient query generator and a robust, plug-and-play extraction pipeline that can significantly improve the success rates of existing LLM-dependent attacks.

\begin{table}[t]
\centering
\caption{Extraction Rate and Query Efficiency results under a 2000-query budget for Agents A, B, and C. Best results are in bold. Higher ER and lower QE are better.}
\label{tab:abc_agents_appendix}
\resizebox{\columnwidth}{!}{%
\begin{tabular}{@{} l cc cc cc @{}}
\toprule
& \multicolumn{2}{c}{\textbf{Agent A}} & \multicolumn{2}{c}{\textbf{Agent B}} & \multicolumn{2}{c}{\textbf{Agent C}} \\
\cmidrule(lr){2-3} \cmidrule(lr){4-5} \cmidrule(lr){6-7}
\textbf{Attack} & ER & QE & ER & QE & ER & QE \\
\midrule
RAGThief+\textsc{snap}             & 44.6\% & 4.48 & 17.4\% & 11.49 & 49.2\% & 4.07 \\
PirateRAG+\textsc{snap}            & 57.4\% & 3.48 & 68.1\% & 2.94 & 26.9\% & 7.43 \\
\textbf{\sys (Ours)} & \textbf{69.7\%}  & \textbf{2.87} & \textbf{68.5\%} & \textbf{2.92} & \textbf{79.3\%} & \textbf{2.52} \\
\bottomrule
\end{tabular}%
}
\end{table}

\section{Computational Cost and API Pricing}
\label{appendix:cost}

We investigate whether the query-efficiency advantage of \sys translates into lower API costs, and what attacker-side computation is required to obtain this advantage. Although \sys avoids using an attacker-side LLM, its query search requires repeated token mutations and embedding evaluations. We therefore compare victim API usage, attacker-side LLM usage, and mean elapsed times to assess the trade-off between API cost and local computation.

\textbf{Evaluation setup.}
We evaluate \sys, RAGThief and PirateRAG across all five datasets, accounting for the work performed until each attack first reaches 50\% extraction. Table~\ref{tab:combined_cost_tariffs} estimates the API charges for different providers, while Table~\ref{tab:combined_cost_metrics} reports the measured workloads and runtimes. Mean elapsed times include attacker preparation, local processing, and waiting for model responses; they exclude prior server and index setup and offline evaluation. We use oracle to denote the attacker-side LLM employed by the competing methods. All cost experiments use two NVIDIA A40 GPUs, with one assigned to the victim and the other to attacker-side computation.

\textbf{API cost estimates.} To determine how the measured workloads affect API charges, we use the recorded input and output token counts, including repeated chunks and unsuccessful queries. Successful near-verbatim extraction requires the victim to receive retrieved chunks in its context and reproduce chunk text in its output. Both input and output usage therefore contribute to the attack cost. Let $I_v,O_v$ and $I_a,O_a$ denote the victim and oracle token counts, respectively. Given their input/output prices $p_{v,i},p_{v,o},p_{a,i},p_{a,o}$ per million tokens, the combined LLM charge is
\begin{equation}
 C_{\mathrm{LLM}} = \frac{
 p_{v,i} I_v + p_{v,o} O_v + p_{a,i} I_a + p_{a,o} O_a}{10^6}.
 \label{eq:measured_token_cost}
\end{equation}

Table~\ref{tab:combined_cost_tariffs} applies different per-token prices to the same recorded workloads. This estimates the effect of pricing while holding the observed query sequences and token counts fixed. It does not establish extraction performance on the listed models, which may differ in tokenization, responses, and latency. We report victim and oracle charges separately because the attacker can choose an oracle independently of the victim, including a locally hosted model with no oracle API fee. For the API prices in Table~\ref{tab:combined_cost_tariffs}, we use OpenAI's
standard uncached short-context prices and Together AI's listed serverless
prices.\footnote{\url{https://www.together.ai/pricing}, October 12, 2026;}\footnote{\url{https://developers.openai.com/api/docs/pricing}, October 12, 2026;}

Even when oracle API fees are excluded, \sys has the lowest victim charge among the reported entries on BioASQ, Neural Bridge, and SciDocs under every listed tariff. It is also the lowest or tied on Wikipedia and TREC-COVID when input and output tokens have equal prices. We observe some exceptions on Wikipedia and TREC-COVID datasets when output tokens are more expensive. At \$2/\$10 per million input/output tokens, \sys costs \$11.56 versus PirateRAG's \$10.98 on Wikipedia, and \$8.43 versus RAGThief's \$7.57 on TREC-COVID. In these cases, \sys sends fewer queries but produces more output tokens, leaving its victim API cost slightly higher. This explains why query efficiency alone does not determine the API bill.

If both the victim and oracle use paid APIs at the same listed tariff, \sys has the lowest combined charge among the reported entries on every dataset. However, its API cost advantage does not depend solely on assuming a paid oracle. Victim-only charges are generally lower and remain comparable in the two exceptions above. 

\textbf{Computation and runtime.} As shown in Table~\ref{tab:combined_cost_metrics}, \sys reaches 50\% extraction with fewer victim calls than each competitor that reaches this threshold on the same dataset. On SciDocs, for example, it requires 214 calls, compared to 466 for RAGThief and 1,614 for PirateRAG. The competing methods also invoke an oracle to generate their queries, whereas \sys has no oracle calls. These results show that \sys reduces interactions with the victim while avoiding the additional LLM generation workload.

We report mean elapsed time per victim call and per generated candidate to account for how the methods prepare their queries. Both averages divide the full measured elapsed time by the corresponding count, including all preparation and model responses. \sys has a higher mean time per victim call on all five datasets. However, it prepares queries in bulk, including candidates that have not yet been submitted. For example, on SciDocs, RAGThief generates 466 candidates for 466 victim calls, while \sys generates 2,426 candidates for 214 calls. If these queued candidates are used later, their existing preparation cost is spread over more victim calls. We do not subtract this cost from the reported victim call time values.

When averaged over generated candidates, \sys has the lowest reported mean elapsed time on Wikipedia, Neural Bridge, TREC-COVID, and SciDocs. On SciDocs, \sys averages 33.30 seconds per candidate, compared to 72.21 for RAGThief and 64.13 for PirateRAG. BioASQ is the exception, where \sys averages 40.21 seconds compared to PirateRAG's 30.68 seconds.

\begin{table*}[p]
\centering
\small
\setlength{\tabcolsep}{2pt}
\renewcommand{\arraystretch}{1.12}
\begin{tabular*}{\textwidth}{@{\extracolsep{\fill}}ll*{10}{r}@{}}
\toprule
Dataset & Method & \multicolumn{2}{c}{\shortstack{Llama 3 8B\\Instruct Lite\\0.14 / 0.14}} & \multicolumn{2}{c}{\shortstack{Llama 3.3\\70B\\1.04 / 1.04}} & \multicolumn{2}{c}{\shortstack{GPT-6\\Luna\\0.10 / 0.50}} & \multicolumn{2}{c}{\shortstack{GPT-6\\Sol\\2.00 / 10.00}} & \multicolumn{2}{c}{\shortstack{GPT-6\\Astra\\10.00 / 50.00}} \\
\cmidrule(lr){3-4}\cmidrule(lr){5-6}\cmidrule(lr){7-8}\cmidrule(lr){9-10}\cmidrule(lr){11-12}
 & & Victim & Oracle & Victim & Oracle & Victim & Oracle & Victim & Oracle & Victim & Oracle \\
\midrule
\multirow{3}{*}{BioASQ} & RAGThief$^{\ddagger}$ & N/A & N/A & N/A & N/A & N/A & N/A & N/A & N/A & N/A & N/A \\
 & PirateRAG & 0.63 & 0.80 & 4.71 & 5.93 & 1.12 & 1.63 & 22.48 & 32.59 & 112.41 & 162.95 \\
 & \sys & 0.32 & 0.00 & 2.40 & 0.00 & 0.60 & 0.00 & 12.07 & 0.00 & 60.34 & 0.00 \\
\addlinespace[4pt]
\multirow{3}{*}{Wikipedia} & RAGThief & 0.30 & 0.55 & 2.20 & 4.05 & 0.64 & 1.72 & 12.73 & 34.43 & 63.64 & 172.14 \\
 & PirateRAG & 0.27 & 1.55 & 2.01 & 11.48 & 0.55 & 3.60 & 10.98 & 72.06 & 54.91 & 360.29 \\
 & \sys & 0.26 & 0.00 & 1.96 & 0.00 & 0.58 & 0.00 & 11.56 & 0.00 & 57.79 & 0.00 \\
\addlinespace[4pt]
\multirow{3}{*}{\shortstack[l]{Neural\\Bridge}} & RAGThief$^{\dagger}$ & 0.52 & 0.56 & 3.88 & 4.16 & 1.17 & 1.71 & 23.32 & 34.14 & 116.58 & 170.72 \\
 & PirateRAG & 0.36 & 1.58 & 2.68 & 11.73 & 0.81 & 3.95 & 16.24 & 79.06 & 81.20 & 395.29 \\
 & \sys & 0.18 & 0.00 & 1.30 & 0.00 & 0.34 & 0.00 & 6.74 & 0.00 & 33.72 & 0.00 \\
\addlinespace[4pt]
\multirow{3}{*}{TREC-COVID} & RAGThief & 0.21 & 0.13 & 1.59 & 1.00 & 0.38 & 0.41 & 7.57 & 8.28 & 37.85 & 41.38 \\
 & PirateRAG & 0.76 & 0.90 & 5.67 & 6.66 & 1.34 & 1.82 & 26.73 & 36.37 & 133.65 & 181.83 \\
 & \sys & 0.21 & 0.00 & 1.55 & 0.00 & 0.42 & 0.00 & 8.43 & 0.00 & 42.17 & 0.00 \\
\addlinespace[4pt]
\multirow{3}{*}{SciDocs} & RAGThief & 0.32 & 0.21 & 2.38 & 1.56 & 0.57 & 0.64 & 11.36 & 12.74 & 56.79 & 63.72 \\
 & PirateRAG & 1.13 & 1.02 & 8.40 & 7.57 & 2.02 & 2.27 & 40.42 & 45.39 & 202.12 & 226.97 \\
 & \sys & 0.17 & 0.00 & 1.24 & 0.00 & 0.33 & 0.00 & 6.69 & 0.00 & 33.45 & 0.00 \\
\bottomrule
\end{tabular*}
\par\smallskip
\begin{minipage}{\textwidth}
\footnotesize\par
$^{\dagger}$ RAGThief stopped at 49.3\% extraction on Neural Bridge; the values include its complete run of 1,135 victim calls.\par
$^{\ddagger}$ N/A: RAGThief did not reach 50\% extraction on BioASQ (maximum 27.9\%).
\end{minipage}

\caption{ Estimated API charges (USD) at the reported extraction endpoints. Input/output rates below each tariff name are in USD per million tokens. Victim and oracle charges are shown separately and can be priced independently.}
\label{tab:combined_cost_tariffs}

\end{table*}

\begin{table*}[p]

\centering
\fontsize{8.5}{10}\selectfont
\setlength{\tabcolsep}{1.5pt}
\renewcommand{\arraystretch}{1.12}
\begin{tabular*}{\textwidth}{@{\extracolsep{\fill}}ll*{10}{r}@{}}
\toprule
Dataset & Method & \shortstack{Victim\\calls} & \multicolumn{2}{c}{\shortstack{Oracle\\generation}} & \shortstack{Query\\candidates} & \multicolumn{2}{c}{\shortstack{Mean elapsed\\time (s)}} & \multicolumn{2}{c}{\shortstack{Victim tokens\\(M)}} & \multicolumn{2}{c}{\shortstack{Oracle tokens\\(M)}} \\
\cmidrule(lr){4-5}\cmidrule(lr){7-8}\cmidrule(lr){9-10}\cmidrule(lr){11-12}
 &  &  & Batches & Prompts &  & Call & Candidate & Input & Output & Input & Output \\
\midrule
\multirow{3}{*}{BioASQ} & RAGThief$^{\ddagger}$ & N/A & N/A & N/A & N/A & N/A & N/A & N/A & N/A & N/A & N/A \\
 & PirateRAG & 1,307 & 2,107 & 3,134 & 1,307 & 30.68 & 30.68 & 2.854 & 1.678 & 3.051 & 2.649 \\
 & \sys & 910 & 0 & 0 & 2,407 & 106.35 & 40.21 & 1.379 & 0.931 & 0.000 & 0.000 \\
\addlinespace[4pt]
\multirow{3}{*}{Wikipedia} & RAGThief & 1,144 & 1,144 & 1,144 & 1,144 & 35.57 & 35.57 & 1.057 & 1.061 & 0.570 & 3.329 \\
 & PirateRAG & 1,610 & 2,487 & 5,813 & 1,610 & 24.43 & 24.43 & 1.049 & 0.888 & 4.788 & 6.248 \\
 & \sys & 953 & 0 & 0 & 3,827 & 94.58 & 23.55 & 0.905 & 0.975 & 0.000 & 0.000 \\
\addlinespace[4pt]
\multirow{3}{*}{\shortstack[l]{Neural\\Bridge}} & RAGThief$^{\dagger}$ & 1,135 & 1,135 & 1,135 & 1,135 & 43.96 & 43.96 & 1.746 & 1.982 & 0.728 & 3.269 \\
 & PirateRAG & 865 & 1,443 & 4,121 & 865 & 44.99 & 44.99 & 1.193 & 1.385 & 4.212 & 7.063 \\
 & \sys & 444 & 0 & 0 & 2,782 & 190.47 & 30.40 & 0.724 & 0.530 & 0.000 & 0.000 \\
\addlinespace[4pt]
\multirow{3}{*}{TREC-COVID} & RAGThief & 260 & 260 & 260 & 260 & 86.99 & 86.99 & 0.962 & 0.565 & 0.169 & 0.794 \\
 & PirateRAG & 941 & 1,216 & 3,604 & 941 & 74.50 & 74.50 & 3.469 & 1.979 & 3.459 & 2.945 \\
 & \sys & 209 & 0 & 0 & 1,849 & 439.06 & 49.63 & 0.814 & 0.681 & 0.000 & 0.000 \\
\addlinespace[4pt]
\multirow{3}{*}{SciDocs} & RAGThief & 466 & 466 & 466 & 466 & 72.21 & 72.21 & 1.438 & 0.848 & 0.282 & 1.218 \\
 & PirateRAG & 1,614 & 1,866 & 3,959 & 1,614 & 64.13 & 64.13 & 5.040 & 3.034 & 3.430 & 3.853 \\
 & \sys & 214 & 0 & 0 & 2,426 & 377.47 & 33.30 & 0.657 & 0.538 & 0.000 & 0.000 \\
\bottomrule
\end{tabular*}
\par\smallskip
\begin{minipage}{\textwidth}
\footnotesize\par
$^{\dagger}$ RAGThief stopped at 49.3\% extraction on Neural Bridge; the values include its complete run of 1,135 victim calls.\par
$^{\ddagger}$ N/A: RAGThief did not reach 50\% extraction on BioASQ (maximum 27.9\%).
\end{minipage}

\caption{ Model usage, mean elapsed times, and token counts for extracting 50\% of each knowledge base, subject to the noted exceptions. We report oracle generation batches and individual prompts separately, since one batch can process multiple prompts. Mean times divide the full elapsed time by victim calls or generated query candidates, respectively. M denotes millions.}
\label{tab:combined_cost_metrics}

\end{table*}

\end{document}